\documentclass[12pt]{article}
\usepackage[margin=1in]{geometry}

\usepackage{times}

\usepackage{amsfonts}%
\usepackage{amssymb}%
\usepackage{amsthm}
\usepackage{bm}
\usepackage{relsize}
\usepackage{graphicx}
\usepackage{caption}
\usepackage{epstopdf}
\usepackage{amsmath}
\usepackage{tikz}
\usetikzlibrary{trees,arrows}
\usetikzlibrary{decorations}
\usetikzlibrary[decorations]
\usepgflibrary{decorations.pathmorphing}
\usepgflibrary[decorations.pathmorphing]
\usetikzlibrary{decorations.pathmorphing}
\usetikzlibrary[decorations.pathmorphing]
\usepackage{booktabs}
\usepackage{natbib}
\usepackage{subcaption}
\usepackage{pseudocode}
\usepackage{pseudocode}
\usepackage{algorithm}
\usepackage{algorithmicx}
\usepackage{algpseudocode}
\usepackage{verbatim} 

\bibpunct{(}{)}{,}{}{}{;} 

\newcommand\independent{\protect\mathpalette{\protect\independenT}{\perp}}
\def\independenT#1#2{\mathrel{\rlap{$#1#2$}\mkern2mu{#1#2}}}

\DeclareMathOperator*{\E}{\mathbb{E}}

\DeclareMathOperator*{\cov}{\text{cov}}

\DeclareMathOperator{\circlearrow}{\hbox{$\circ$}\kern-1.5pt\hbox{$\rightarrow$}}
\DeclareMathOperator{\circlecircle}{\hbox{$\circ$}\kern-1.2pt\hbox{$--$}\kern-1.5pt\hbox{$\circ$}}
\DeclareMathOperator{\starstar}{\hbox{$\ast$}\kern-1.1pt\hbox{$--$}\kern-1.1pt\hbox{$\ast$}}
\DeclareMathOperator{\rightarrowstar}{\hbox{$\ast$}\kern-1.5pt\hbox{$\rightarrow$}}
\DeclareMathOperator{\leftarrowstar}{\hbox{$\leftarrow$}\kern-1.5pt\hbox{$\ast$}}

\DeclareMathOperator*{\Pa}{Pa}
\DeclareMathOperator*{\Ne}{Ne}
\DeclareMathOperator*{\Adj}{Adj}
\DeclareMathOperator*{\An}{An}
\DeclareMathOperator*{\De}{De}
\DeclareMathOperator*{\G}{\mathcal{G}}

\DeclareMathOperator*{\sk}{sk}
\DeclareMathOperator*{\pds}{\text{pds}}

\newtheorem{Thm}{Theorem}

\title{\bf A cautious approach to constraint-based causal model selection}

\author{Daniel Malinsky\footnote{Corresponding author: d.malinsky@columbia.edu} \\
Department of Biostatistics\\
Columbia University
}

\date{Draft: August 16, 2026}

\begin{document}

\maketitle

\begin{abstract}
	We study the data-driven selection of causal graphical models using constraint-based algorithms, which determine the existence or non-existence of edges (causal connections) in a graph based on testing a series of conditional independence hypotheses. In settings where the ultimate scientific goal is to use the selected graph to inform estimation of some causal effect of interest (e.g., by selecting a valid and sufficient set of adjustment variables), we argue that a ``cautious'' approach to graph selection should control the probability of falsely removing edges and prefer dense, rather than sparse, graphs. We propose a simple inversion of the usual conditional independence testing procedure: to remove an edge, test the null hypothesis of conditional association greater than some user-specified threshold, rather than the null of independence. This equivalence testing formulation to testing independence constraints leads to a procedure with desriable statistical properties and behaviors that better match the inferential goals of certain scientific studies, for example observational epidemiological studies that aim to estimate causal effects in the face of causal model uncertainty. We illustrate our approach on a data example from environmental epidemiology.
\end{abstract}

\section{Introduction}

Causal graphical models are used in many scientific domains (especially the biological, health, and social sciences) to represent important causal assumptions about the 
processes that underlie collected data, e.g.\ assumptions about study design, relevant variables, sources of confounding or other potential biases. When a causal graphical model is known or assumed to be correct in a given analytical setting, it may be used for various purposes: to establish identifiability (or non-identifiability) of target quantities, to aid in the estimation of causal effects (via formal adjustment criteria or other algorithms), to draw qualitative causal conclusions about causal pathways or network topology, or to guide reasoning about generalizability or mechanistic explanation. There is an abundant theoretical literature on how to reason about causal identification and how to formulate efficient estimators of effects in the context of a given graphical model: for an overview of identification results see \citet{shpitser2018identification}; for some recent work on exploiting graphical structure for efficient estimation see \citet{rotnitzky2020efficient,bhattacharya2022semiparametric,smucler2022efficient} and \citet{guo2023variable}.
Graphical structure learning algorithms, which estimate graphs from data, have also been extensively studied in computer science and statistics. Such algorithms are regularly applied to answer substantive research questions, especially high-dimensional biological applications (e.g., genetics and neuroscience) where most connections are not known a priori \citep{sgs2000, stekhoven2012causal,  sanchez2019estimating, dubois2020causal, cai2022causal, chen2022individualized}.

The focus of this work is on graphical structure learning for the ``downstream'' purpose of using the estimated graph for a subsequent causal inference tasks, such as establishing the identifying formula for some causal effect of interest and then estimating it. In particular, our interest is in ``moderate-dimensional'' settings (i.e., more than 4 variables but fewer than hundreds) where the causal graph is not known, or perhaps only partially known, and sparsity should not be assumed. This is arguably the setting of many epidemiologic studies, where it is common to specify a directed acyclic graph with ~5-30 vertices on the basis of domain knowledge and then use this graph to select a valid adjustment set to estimate some effect. \citet{tennant2021use} document hundreds of papers in the applied health sciences that conform to this basic setup, almost all of which posit a causal graph and then use regression with a graph-determined adjustment set in their final estimate of some causal effect. Interestingly, the majority of published graphs assessed in this review are dense but not completely connected: they are missing edges that correspond to conditional independencies (d-separation relations, a.k.a.\ ``exclusion restrictions'') in the assumed statistical model. Though these structures have empirically testable implications, they are rarely tested against the data in practice \citep{tennant2021use, ankan2023simple} and the graphs are even less commonly selected by structure learning algorithms in this domain. It is plausible that some of the posited graphical models may be rejected on the basis of the data if put to test, or that substantively different models would be selected in a data-driven manner by structure learning algorithms.

An important obstacle to having confidence in existing causal structure learning procedures in such applied health science settings is that they have a tendency to estimate structures that are overly sparse, i.e., missing too many edges compared to the ``ground truth.'' Consider, in particular, classical constraint-based structure learning algorithms such as the PC (``Peter-Clark") algorithm, the FCI (``Fast Causal Inference") algorithm, or variants thereof \citep{sgs2000}. These algorithms, described in more detail below, proceed by testing a sequence of conditional independence hypotheses (constraints) and thereby narrow down a graph or set of graphs that are compatible with these test results. Though much statistical theory has been dedicated to controlling false positives in the sense of false edge inclusions (multiple testing adjustments, limits on the false discovery rate), it is relatively ``easy'' in practice to achieve very few false edge inclusions with standard algorithms \citep{drton2007multiple,drton2008sinful,strobl2019estimating}. Indeed, even some of the most basic algorithms such as PC, with appropriate setting of the tuning parameter, will achieve $<10\%$ false edge inclusions (high precision) in simulation studies \citep{raghu2018comparison}. However, it is difficult to achieve low rates of error for false edge exclusions (high recall). This is problematic for downstream causal effect estimation tasks, where bias is largely driven by high rates of wrongly excluded edges. This will be made more precise below.

This paper proposes to reformulate the conditional independence hypothesis tests of classical constraint-based algorithms as equivalence tests: test the null hypothesis of association greater than some (user-chosen, sample-size dependent) threshold $\delta > 0$, rather than test the null of no association. We argue that this addresses three inter-related issues in applied causal model selection: 1) that it is empirically ``easy'' to tune algorithms for high precision but ``hard'' to tune algorithms for high recall of edges; 2) removing edges from a dense starting graph whenever the null hypothesis of independence fails to be rejected amounts to a statistical decision with poor error control, namely ``accepting the null'' repeatedly in tests of possibly low power; and 3) statistical ``caution'' (or ``conservatism") would err on the side of more dense graphs rather than more sparse graphs, since estimating causal effects in a supermodel of the truth (a larger set of distributions that includes the true distribution) preserves consistency of estimators, whereas inference in a sparse submodel is liable to introduce bias if the submodel is incorrect. Our focus is on constraint-based algorithms based on hypothesis tests primarily because these algorithms do not need to rely on strong parametric assumptions: they enjoy statistical guarantees even in nonparametric settings with the incorporation of appropriate nonparametric tests of association. Furthermore, constraint-based algorithms are somewhat simpler to explain from a statistical perspective and more transparent from a user perspective (since each algorithmic decision can be traced to an independence test or one of a few graphical rules). We draw on the existing statistical theory of consistency for constraint-based algorithms. Investigation of related ideas in alternative approaches to structure learning (e.g., score-based approaches) are left to future work. 

Our contributions are the following. First, we propose an informal but useful distinction between two broadly different research ``contexts'' for causal structure learning, characterized by different scientific and statistical goals. One context is the \emph{context of causal discovery}, wherein the focus is on confidently asserting the existence of edges, usually edges that are ``strong'' in some sense, perhaps promising for the planning of future experiments or interventions. In this context, there is thought to be a substantial cost to falsely asserting the existence of a causal edge. Alternatively, a distinct set of goals comprises what we call the \emph{context of submodel selection}, wherein the aim is to select a (causal and statistical) submodel of the maximally agnostic saturated model that imposes no independence constraints on the data distribution. That is, the starting point is presumed to be the set of all possible distributions over the relevant variable set and so selecting any edge-subgraph of the complete graph is selecting a submodel, which then may be the analysis model that forms the basis of subsequent inferences. In this context, asserting the non-existence of an edge is the more risky decision.
Next, we propose that in the goals of accurate submodel selection can be better achieved by reversing the usual null hypothesis testing approach of constraint-based algorithms: instead of treating the conditional independence hypothesis as null for every pair of variables and removing an edge when the null fails to be rejected, one may test a null corresponding to dependence (association) greater than some tolerance level and only remove an edge when the estimated association is tolerably small (i.e., when the null of substantial dependence is rejected). We implement the proposal in the multivariate Gaussian setting and describe how tests may be formulated with other more general measures of association. We show in simulations that performance is improved in a specific sense: recall can be controlled to a high level and the estimated graph has a density closer to the density of the true graph. 
We discuss some theoretical properties that attend to reformulating the algorithms this way and how to select the tolerance parameter. Finally, we apply the proposed approach to a causal effect question in observational environmental epidemiology.

\section{Background}

Let $X = (X_1,...,X_p)$ be a random vector. We may equivalently write $X = (X_i)_{i \in V}$ to denote a random vector with elements indexed by the finite set $V$ (e.g., $V= \{1,...,p\}$). $X$ may take values in a continuous or discrete space. We assume this random vector is defined with respect to a probability space $(\Omega, \mathcal{A}, P)$ and the measure $P$ has density $p(x)$ with respect to some dominating measure (e.g., Lebesgue or counting measure). We will assume that the variables are all continuous or all discrete, though mixtures of continuous and discrete variables may be accomodated with some additional effort.
We use the shorthand $X_S$ to denote the subvector $(X_i)_{i \in S}$ where $S \subseteq V$. 
A model $\mathcal{M}$ is a set of densities and an independence model is a set of densities satisfying some conditional independence constraints. 

\subsection{Graphs}

A graph $\G = (V,E)$ is pair where $V$ is a finite set of \emph{vertices}, e.g., $V = \{1,...,p\}$, and $E$ is a set of ordered pairs of vertices 
called \emph{edges}. An edge $(i,j) \in E$ with also $(j,i) \in E$ is called \emph{undirected} whereas an edge $(i,j) \in E$ with $(j,i) \notin E$ is called \emph{directed}. We denote an undirected edge by $i-j$ and a directed edge by $i \rightarrow j$. (More general graphs with additional edge types can be defined with some additional notation, e.g., it is also possible to define mixed graphs with bidirected $\leftrightarrow$ edges and others.) If $i \rightarrow j$ then $i$ is a \emph{parent} of $j$ and we denote the set of parents of a vertex $j$ in graph $\G$ by $\Pa_j(\G)$ or $\Pa_j$ when the relevant graph is clear. If $i - j$ then $i$ is a \emph{neighbor} of $j$ and we denote the set of neighbors of a vertex $j$ in graph $\G$ by $\Ne_j(\G)$ or $\Ne_j$. If there is an edge between two vertices in $\G$ we say they are \emph{adjacent} and denote the adjacency set of $j$ in $\G$ by $\Adj_j(\G)$ or $\Adj_j$. A \emph{path} $\pi$ is a finite sequence of distinct vertices such that consecutive vertices $(i,i+1)$ in the sequence are adjacent. A \emph{directed path} from $j$ to $k$ is a path beginning at $j$ and ending at $k$ such that consecutive vertices $(i,i+1)$ are joined by a directed edge from $i$ to $i+1$. A \emph{directed cycle} is a directed path from some vertex to itself. A directed acyclic graph (DAG) is a graph with only directed edges and no directed cycles. An undirected graph (UG) is a graph with only undirected edges. A graph is called \emph{mixed} if it has more than one kind of edge, e.g., a graph with both directed and undirected edges (and possibly bidirected edges, discussed below). In a directed graph, if there is a directed path from $i$ to $j$ then $i$ is an \emph{ancestor} of $j$ and $j$ is a \emph{descendent} of $i$. Denote the sets of ancestors of $j$ by $\An_j(\G)$ and descendents by $\De_j(\G)$. By convention every vertex is also counted as its own ancestor and descendent. 

We will write $\G \subseteq \G'$ when $\G = (V,E)$ is an edge-subgraph of $\G' = (V,E')$, i.e., the two graphs share the same vertices and $E \subseteq E'$ (the set of edges $E$ in $\G$ is a subset of the edges $E'$ in $\G'$). $\G'$ is then called a supergraph of $\G$. The skeleton of a graph $\G$ is the graph formed by replacing all edges in $\G$ with undirected edges. We denote the skeleton of $\G$ by $\sk(\G)$ and write $\sk(\G) \subseteq \sk(\G')$ when the subgraph relation holds ignoring edge directions.

Some additional basic graphical definitions follow. Given a path $\pi$ in a graph $\G$, a non-endpoint vertex $k$ on $\pi$ is called a collider if the two edges incident to $k$ are both into $k$, i.e., have arrowheads at $k$ as in $i \rightarrow k \leftarrow j$. If the colliding vertices (here $i,j$) are not adjacent (not directly connected by an edge) the substructure is called an unshielded collider. A path $\pi$ in $\G$ between distinct vertices $i$ and $j$ is called a d-connecting path conditional on vertex set ${C}$ (${C} \subseteq V \setminus \{i, j\}$) if every collider on $\pi$ is in ${C}$ or is in $\An_C(\G)$ and every non-collider on $\pi$ is not in ${C}$. $i$ and $j$ are \emph{d-separated} or \emph{blocked} conditional on ${C}$ (written: $i \perp^{\G}_d j \mid {C}$) if there is no d-connecting path conditional on ${C}$ between $i$ and $j$. ${A} \perp^{\G}_d {B} \mid {C}$ if $i \perp^{\G}_d j \mid {C}$ for all $i \in {A}$ and $j \in {B}$.

\subsection{Graphical Independence Models}

We will identify the elements of random vector $X = (X_i)_{i \in V}$ with the vertices $V$ of a DAG $\G$. The distribution $p(x)$ is said to factorize wrt to DAG $\G$ if $p(x) = \prod_{i=1}^{p} p(x_i \mid x_{\Pa_i(\G)})$, i.e., the joint factorizes into a product of conditional distributions, one for each node given its parents in $\G$. A distribution is said to satisfy the global Markov property wrt DAG $\G$ if for disjoint subsets $A, B, C \subset V$, $A \perp_d^{\G} B \mid C \implies X_A \independent X_B \mid X_C$, i.e., d-separation in the graph $\G$ implies conditional independence in $p(x)$. A distribution is said to satisfy the local Markov property wrt DAG $\G$ if every variable is independent of its non-descendents given its parents in $\G$. A fundamental result in the theory of directed acyclic graphical models is that the factorization, global Markov, and local Markov properties are all equivalent \citep{lauritzen1996graphical}, so we can generically say a distribution $p(x)$ satisfies ``the Markov condition'' wrt DAG $\G$ when it satisfies any of these. 

Graphical models are conditional independence models, i.e., they represent sets of distributions that satisfy conditional independence constraints. Given a DAG $\G$, it is easy to ``read off'' the conditional independence constraints encoded by $\G$ using the above Markov properties, most commonly using the d-separation criterion.

Multiple graphs may imply the same conditional independence constraints, in which case we say they are \emph{Markov equivalent}. 
There is a simple criterion for determining whether two DAGs are Markov equivalent: $\G_1$ and $\G_2$ are Markov equivalent if and only if they share the same adjacencies and unshielded colliders \citep{verma1990equivalence}. The set of graphs that imply the same set of conditional independence constraints form a Markov equivalence class. The Markov equivalence class of a DAG $\G$ may be represented by a single mixed graph called a completed partially directed acyclic graph (CPDAG), which may contain both directed and undirected edges.  Formally, the CPDAG of a DAG $\G$, denoted by $\G'$, is a mixed graph that has the same skeleton as $\G$ where a directed edge occurs in $\G'$ if and only if it appears in all DAGs Markov equivalent to $\G$.

\subsection{Causal Graphical Models}

Causal graphical models are graphical models endowed with some causal interpretation. One common approach to mathematically formulating a causal interpretation is to posit a system of (nonparametric) structural equations, also called a structural causal model (SCM), associated with a DAG $\G$ \citep{pearl2009causality, peters2017elements}. This is a system of equations wherein each variable corresponding to a vertex on the DAG $\G$, is a function of its graphical parents and independent error: $X_j = f_j(X_{\Pa_j(\G)}, \epsilon_j)$ $\forall j \in V$. (Here $X_{\Pa_j(\G)}$ to refers to the subset of $X$ in $\Pa_j(\G)$.) It is typical to assume that the errors are mutually independent, i.e., the distribution of errors factorizes: $p(\epsilon_1,...,\epsilon_p) = \prod_{j=1}^p p(\epsilon_j)$. An intervention that sets some variable $X_i$ to value $x_i$ is represented by a modified system of equations wherein the equation for $X_i$ is replaced by the constant equation $X_i = x_i$ and the value $x_i$ replaces $X_i$ in any remaining equations where $X_i$ appears as an argument. The new joint distribution over vector $X$ induced by this modified system of equations is called the post-intervention distribution. More details about the semantics and properties of SCMs can be found in \citet{pearl2009causality}. 

Another common approach to associating a precise causal interpretation with a DAG begins with positing a collection of potential outcome random variables. $X_j(x_i)$ generically denotes the value of $X_j$ had the variable $X_i$ been set, possibly contrary to fact, to value $x_i$. Potential outcome (a.k.a.\ counterfactual) variables such as these are the fundamental ingredients used to define various causal effect parameters of interest, for example average treatment effects (ATEs) \citep{hernan2020causal}. In a multivariable graphical setting, one may begin by assuming the existence of all ``one-step-ahead'' potential outcomes of the form  $X_j(x_{\Pa_j})$, i.e., the value of $X_j$ had its parents $\Pa_j(\G)$ been set to values $x_{\Pa_j}$. Given the set of one-step-ahead potential outcomes for all $j \in V$, for any $A \subseteq V$ the potential outcome $X_j(a)$ had $X_A$ been set to $a$ is defined by recursive substitution $X_j(a) \equiv X_j(a \cap x_{\Pa_j}, \{X_k(a) \mid k \in \Pa_j \setminus A \})$. Causal models impose independence restrictions on the set of potential outcomes. For example, one influential model due to Robins asserts that the variables $\{ X_j(x_{\Pa_j}) \mid j \in V \}$ are mutually independent for every fixed vector of assignments $x$ to $X$ \citep{robins1986new}. Another (stronger) set of independence assumptions is equivalent to the independence of errors $\epsilon_j$ in the SCM above \citep{robins2010alternative}. In either case, the post-intervention joint distribution for any $X(a)$ derived from one-step-ahead potential outcomes and recursive substitution is identified via the following formula: $ p(X(a)) = \prod_{j=1}^{p} p(X_j \mid X_{\Pa_j \setminus A}, a \cap x_{\Pa_j}) $ provided a positivity condition holds. This factorization is called the \emph{extended g-formula}. Additional connections between the set of posited potential outcomes and graphical concepts may be bridged by introducing a certain graph derived from $\G$ called a Single World Intervention Graph (SWIG); see \citet{richardson2013single}. 
One may also formulate an extension of Pearl's celebrated ``do-calculus'' for deriving post-intervention distributions using potential outcomes in combination with SWIGs \citep{malinsky2019potential}.

There are various conceptual elucidations and translations that may be drawn between SCMs and potential outcome models \citep{richardson2013single, peters2017elements, malinsky2019potential}. There are also presentations that instead foreground proposed causal principles related to ``invariance'' or asymmetries of information \citep{scholkopf2021toward}. For the purposes of the graphical structure learning ideas that are the focus of this paper, it does not make any difference which formalism underlies the intended causal interpretation of a DAG; this is because the structure learning algorithms we describe here only select on the basis of observable independence constraints (among ``factual,'' not potential outcome variables) implied by a given DAG $\G$, on which the different causal models for $\G$ agree. So, one may have either SCMs or potential outcome models in mind when using structure learning algorithms and we do not elaborate further on causal formalisms here.

Though many theoretical discussions and basic structure learning algorithms focus on DAGs where all variables are observed (i.e., vertices correspond to observable variables), it is often more appropriate to assume that observables may be impacted by unobserved (latent) variables. In this case, the analyst may assume the data is generated according to a DAG $\G$ with latent variables, with vertices $V \cup L$ where indices in $V$ correspond to observed variables and indices in $L$ correspond to unobserved variables. In that case, the target of structure learning will often be a summary graph defined by \emph{latent projection} of the $\G$ onto only observables $(X_i)_{i \in V}$. One such graphical representation the maximal ancestral graph (MAG) \citep{richardson2002ancestral}. A MAG is a mixed graph that may contain both directed ($\rightarrow$) and bidrected ($\leftrightarrow$) edges. A bidirected edge between vertices $i$ and $j$ in a MAG indicates that $i$ and $j$ share some latent parent in the underlying DAG $\G$. MAGs thus represent causal relationships that obtain among observed variables when an arbitrary number and arrangement of latent variables have been marginalized out. MAGs are constructed to satisfy an ``ancestrality'' constraint that rules out directed cycles and almost directed cycles, where an almost directed cycle is a directed path from $i$ to $j$ when $i \leftrightarrow j$. They are also ``maximal'' in the sense that each non-adjacency in a MAG corresponds to some conditional independence constraint among variables in $(X_i)_{i \in V}$. (In their full generality, MAGs may also contain undirected edges to represent associations induced by selection bias; to simplify matters in this presentation we exclude the possibility of selection bias and undirected edges in the MAG. Technically, what we describe here is a subset of MAGs called \emph{directed} MAGs.) Conditional independence constraints implied by a MAG model may be derived from the graph using a simple extension of d-separation called \emph{m-separation}. Similarly to DAGs, multiple MAGs may imply the same conditional independence constraints -- the corresponding Markov equivalence class of a MAG is represented by summary graph called a partial ancestral graph (PAG). PAGs may contain the following edge types: $\rightarrow, \leftrightarrow, \circlearrow,$ and $\circlecircle$, where the $\circ$-mark represents uncertainty about the presence of a tail or arrowhead. (If selection bias is not ruled out, PAGs may also contain undirected or partially undirected edges.) Many structure learning algorithms designed for the setting with possible unmeasured confounding aim to estimate the structure of a PAG \citep{zhang2008causal, zhang2008completeness, colombo2012learning}.

\section{Structure Learning}

Structure learning algorithms estimate the connections in graphical models from observational data. It is typical to begin with the assumption that the observed data is a random sample from some distribution $p(x)$ that satisfies the Markov condition wrt ``true'' DAG $\G_0$. It is also common to further assume that $p(x)$ satisfies the faithfulness assumption wrt $\G_0$: if  $ X_A \independent X_B \mid X_C $ then $A \perp_d^{\G_0} B \mid C$ ($A, B, C \subset V$). Faithfulness, which rules out so-called ``accidental'' or ``non-graphical'' conditional independencies, is controversial and can be violated in practice; see discussion in \citet{uhler2013geometry} and \cite{zhang2016three}. Algorithms have been proposed that enjoy consistency guarantees under conditions weaker than faithfulness \citep{spirtes2014uniformly, solus2021consistency}. For simplicity, we focus in the following discussion on a class of algorithms that typically do require faithfulness, though the proposal may have implications for algorithms outside this class, as we return to later.

\subsection{Constraint-based Model Selection}

Constraint-based structure learning algorithms estimate a graph based on directly testing conditional independence constraints, and select a graphical model that is consistent with all ``discovered'' independence constraints. A ``classical'' algorithm for constraint-based model selection is the PC algorithm \citep{sgs2000}. PC estimates a CPDAG based on hypothesis tests of conditional independence. The closely related FCI algorithm executes many of the same steps and makes use of similar graphical results, but estimates a PAG thus allowing for the possibility of latent confounding \citep{zhang2008completeness}. Though arguably most observational studies should allow for this possibility of latent confounding and aim to estimate PAGs, the PC algorithm is easier to analyze and understand, so we focus primarily on PC in our exposition and consider FCI in Section 8. (FCI is algorithmically very similar to PC, so our discussion of PC will carry over to FCI directly.) Pseudocode for PC is reproduced as Algorithm 3.1.
\begin{figure*}
	\begin{pseudocode}[ruled]{PC}{\textsc{Test}, \alpha}
		\mbox{\textbf{Input:} Samples of the vector $X = (X_1,...,X_p)$} \\
		\mbox{\textbf{Output:} CPDAG $\G$} \\
		\mathbf{1.} \hspace{1.5 mm} \mbox{Form the complete graph $\G$ on vertex set $V = \{1,...,p\}$ with undirected ($-$) edges.} \\
		\mathbf{2.} \hspace{1.5 mm} \mbox{Let } s = 0 \\
		\mathbf{3.} \hspace{1.5 mm} \mbox{\bf{repeat}}\\
		\mathbf{4.} \hspace{1.5 mm} \hspace{5 mm}\FORALL \mbox{pairs of adjacent vertices $(i,j)$ s.t. $|\Adj_i(\G)\setminus \{j\}| \geq s$}\\ 
		\hspace{9.5 mm} \mbox{and subsets $S \subseteq \Adj_i(\G)\setminus \{j\}$ s.t. $|S|=s$}\\
		\mathbf{5.} \hspace{1.5 mm} \hspace{10 mm} \IF \mbox{$X_i \independent X_j \mid X_S$ according to (\textsc{Test}, $\alpha$)}\\ \hspace{15 mm} \mbox{\textbf{then}} \BEGIN \mbox{Delete edge $i - j$ from $\G.$}\\
		\mbox{Let sepset$(i,j) =$ sepset$(j,i) = S$.} \END \\
		\mathbf{6.} \hspace{1.5 mm} \hspace{5 mm} \mbox{\bf{end}}\\
		\mathbf{7.} \hspace{1.5 mm} \hspace{5 mm} \mbox{Let } s = s+1\\
		\mathbf{8.} \hspace{1.5 mm} \mbox{\textbf{until} for each pair of adjacent vertices $(i,j)$, $|\Adj_i(\G)\setminus \{j\}|<s$.}\\
		\mathbf{9.} \hspace{1.5 mm} \FORALL \mbox{triples $(i,k,j)$ s.t. $i \in \Adj_k(\G)$ and $j \in \Adj_k(\G)$}\\ 
		\hspace{5 mm} \mbox{but $i \not\in \Adj_j(\G)$, orient $i \rightarrow k \leftarrow j$ in $\G$ iff $k \not\in \mbox{sepset}(i,j)$.}\\
		\mathbf{10.} \hspace{1.5 mm} \mbox{Exhaustively apply orientation rules (R1-R4) to orient}\\ \hspace{6.5 mm} \mbox{remaining undirected edges.}\\
		\mathbf{11.} \hspace{1.5 mm} \mbox{\textbf{return} $\G$}.
	\end{pseudocode}\\
	\noindent \textbf{Orientation Rules:}
	\begin{itemize}
		\item[R1:] Orient $j - k$ into $j \rightarrow k$ whenever there is an arrow $i \rightarrow j$ such that $i$ and $k$ are nonadjacent.
		\item[R2:] Orient $i - j$ into $i \rightarrow j$ whenever there is a path $i \rightarrow k \rightarrow j$.
		\item[R3:] Orient $i - j$ into $i \rightarrow j$ whenever there are two paths $i-k \rightarrow j$ and $i - l \rightarrow j$ such that $k$ and $l$ are nonadjacent.
		\item[R4:] Orient $i - j$ into $i \rightarrow j$ whenever there are two paths $i - k \rightarrow j$ and $i - l \rightarrow k$ such that $j$ and $l$ are nonadjacent.
	\end{itemize}
\end{figure*}

The PC algorithm proceeds in two stages. In the first stage, the algorithm begins with a complete undirected graph and executes a sequence of conditional independence tests, with conditioning sets of increasing size. If a conditional independence between $i$ and $j$ is detected, the edge between $i$ and $j$ is removed from the graph. The second stage orients edges, first using the ``collider rule'' and then using a series of additional rules. The collider rule follows from the assumptions of faithfulness and acyclicity of the graph: an unshielded triple $(i,j,k)$ must be oriented as $i \rightarrow k \leftarrow j$ iff $k$ is not a member of a d-separating set for $i,j$. The rules R1-R4 simply enforce the acyclicity assumption and prevent any additional unshielded colliders that are not determined by the collider rule. These orientation steps, including the rules R1-R4, have been proven to be \emph{sound} and \emph{complete} in the sense that they all follow from the assumptions and no additional orientations may be determined without additional information \citep{sgs2000}. (R4 is not actually needed for completeness except in the setting where PC is used with background knowledge; see \citet{meek1995causal}.)

The first stage of the PC algorithm requires specifying a test of conditional independence and is modular in the sense that any consistent test may be used here. The standard implementation of this step is formulated as a test of the null hypothesis of independence against a universal alternative of dependence. For example, using Pearson's partial correlation coefficient $\rho$ as the chosen measure of dependence, the null is $\rho_{ij.S}=0$ for each combintation of $i,j,$ and $S$ and the algorithm removes an edge if this null is rejected according to a standard level-$\alpha$ test for user-specified $\alpha$. 

It has become standard to modify the traditional PC algorithm to be ``order-independent'' (insensitive to the ordering of the tests within each iteration of the for-loop) by a variation called PC-stable \citep{colombo2014order}. PC-stable modifies the for-loop in lines 4-8 of Algorithm 3.1 in the following way: at the beginning of the for-loop, define for each $i$ a set $\Adj^*_i = \Adj_i(\G)$. This is a global variable that is updated only once at the beginning of each iteration of the entire loop (i.e., between lines 3 and 4). Then, replace each instance of $\Adj_i(\G)$ in the loop with $\Adj^*_i$ so edge deletions in line 5 no longer affect which other conditional independencies are checked for other pairs of variables at this same level of $s$. We will use the PC-stable modification throughout in the following.
Other modifications and variants of the PC algorithm have been proposed, but they share the same algorithmic ingredients just described \citep{gretton2009nonlinear, harris2013pc,sondhi2019reduced,chakraborty2022nonparametric}. One design feature of the PC algorithm that is worth noting is that, in practice with finite samples, conflicting conditional independence test results and naive application of the orientation rules may lead the algorithm to produce a graph that is not a valid CPDAG. Different implementations of PC or variations of the algorithm may deal with conflicting test results/orientations in distinct ways: some versions allow for bidirected edges or cycles in the final output when conflicts occur and other versions prohibit these via some ``conflict resolving'' approach that may force the output to be a valid CPDAG but may have other drawbacks \citep{ramsey2006adjacency,hyttinen2014constraint,andrews2021practical}.

\subsection{Measuring performance}

Measuring the performance or accuracy of structure learning algorithms in finite samples can be subtle. In contrast to most standard parameter estimation problems in statistics, the estimand is a discrete object (not a point on the real line or an interval), and so there is no one ``standard'' metric of success akin to mean squared error. Let $\G_0$ denote the true graph (i.e., the CPDAG representation of the DAG that generated the data) and $\widehat{\G}$ the estimated graph produced by PC or some learning algorithm. One way to quantify the ``distance'' between $\G_0$ and $\widehat{\G}$ is the structural Hamming distance (SHD) which counts the number of edge additions, deletions, and reversals required to transform $\widehat{\G}$ into $\G_0$. A lower SHD indicates the estimated graph is nearer to the truth, and this can be used as a metric to evaluate or compare performance of different algorithms. However, large differences in SHD can come about in a number of different ways, and so this measure is not very informative about the behavior of learning algorithm: is it missing too many adjacencies? Producing too many extraneous edges? Incorrectly orienting edge directions? For this reason is often desirable to look at multiple components of accuracy that provide more fine-grained detail about what is correct and what is incorrect about the estimated graph.

For every pair of vertices $i,j \in V$, $\widehat{\G}$ may have an adjacency between $i,j$ or it may not. It is common to borrow terminology from the evaluation of classification tasks: if $i,j$ are adjacent in $\widehat{\G}$ but not in $\G_0$, this is called a ``false positive'' (FP) and if $i,j$ are not adjacent in $\widehat{\G}$ but are adjacent in $\G_0$, this is a ``false negative'' (FN). Likewise, if the estimated graph and true graph agree on an adjacency, this is a ``true positive'' (TP) and if they agree on a non-adjacency, this is a ``true negative'' (TN). Precision, defined as the total number of true positives as a fraction of all positives asserted (TP/(TP+FP)), quantifies how accurately the estimator is when it asserts adjacencies. Recall, defined as the fraction of true adjacencies that were asserted as adjacent (TP/(TP+FN)), quantifies how many of the non-adjacencies are correctly asserted by the estimator. Precision is equal to one minus the false discovery rate (FDR) and recall (also called sensitivity or true positive rate) is one minus the false negative rate (FNR). A precision value of 1 indicates that all adjacencies in the estimated graph are correct (there are no false adjacencies) and a recall of 1 asserts that all non-adjacencies are correct (there are no falsely missing edges). 

The same concepts can be extended to orientations: e.g., if there is an arrowhead at $j$ from $i$ in $\widehat{\G}$ but not $\G_0$ this is an arrowhead-FP, etc., and we can define arrowhead precision, arrowhead recall, tail precision, tail recall, and so on. Since the focus of the present discussion will be mostly on adjacencies, precision and recall can be assumed hereafter to refer to adjacency precision and recall.

As already mentioned, typically classical constraint-based structure learning algorithms can be tuned to have quite high precision for some choice of $\alpha$, but thereby have typically poor recall. That is, by setting $\alpha$ to be small, the algorithm requires very strong ``evidence'' against the null hypothesis of conditional independence for each test in order to retain an edge. Small settings of $\alpha$ (e.g., $\alpha = 0.01, 0.001, 0.0001$ or smaller) are both common in research papers that apply or evaluate these algorithms,\footnote{Choices of $\alpha$ have been justified with reference to simulation studies, though of course the performance is dependent on the specific simulation settings and choice of evaluation metric. For example, \citet{kalisch2007estimating} find in their simulations with PC that average SHD is minimized in the range of $\alpha$ between 0.01 and 0.005; also \citet{raghu2018comparison}} and from the theoretical perspective $\alpha_n \to 0$ as $n \to \infty$ is a condition in consistency proofs \citep{kalisch2007estimating}. Empirically, it is typical to observe in simulation studies or real data applications (where ground truth is at least partially known) that ``strong'' edges are likely to be retained (or rather, edges that correspond to dependencies with low observed p-values for the sequence of calculated test statistics) and many real edges are likely to be removed. One partial explanation is that higher-cardinality conditional independence tests -- remember that PC begins with marginal independence tests, and then conditions on subsets of increasing cardinality -- will have low power for detecting associations, and so failing to reject the null will be a frequent occurance. Empirically, it is typical that very few of the retained edges will be false positives, but many of edges removed may be false negatives. This is especially acute when the true graph is dense, not sparse. This is related to the theoretical observation that in dense graphs, near violations of faithfulness may be common, in the sense that nearly unfaithful distributions have large measure \citep{uhler2013geometry}. 

Though recall can be somewhat improved by using higher values of $\alpha$ (though the observed behavior is somewhat complicated, see below), fundamentally there is a tradeoff between precision and recall since tuning $\alpha$ only directly controls the frequency of Type I errors in the standard null hypothesis testing setup. 

\subsection{Two contexts for learning}

If there is a tradeoff between precision and recall, which aspect of performance should be prioritized in practice? There are at least two distinct research contexts to consider: the context of causal discovery and the context of submodel selection. In the context of discovery, the scientific goal served by structure learning is to identify ``strong,'' novel, and/or promising causal relationships from among many possible but uncertain relationships, in order to narrow subsequent research focus, plan possible experiments, or distinguish the most promising targets of intervention from dead ends. The paradigmatic example 
is high-dimensional genetics research, wherein the goal is often to identify a small number of important genes that are promising targets for interventions or that should be prioritized in planning experiments (knowing it is infeasible to carry out all possible experiments over thousands of possible targets and combinations). In the context of discovery, the true (important) causal relationships are often assumed to be sparse, and so the learning problem is analogous to discriminating useful ``signal'' from noisy ``background.''
Exemplary work in this domain has been carried out using causal discovery algorithms, often in conjunction with various improvements or modifications and additional techniques \citep{maathuis2010predicting,stekhoven2012causal,engelmann2015causal,belyaeva2021causal,chen2022individualized}.

In the context of submodel selection, there is one or a small number of causal relationships of specific interest, set against a collection of possible models that imply different statistical strategies for (e.g.)\ identification or estimation of a target quantity. For example, it might be that the ultimate scientific goal is to quantify the effect of one exposure $A$ on some outcome $Y$, against the background of a statistical model for the distribution of $X = (W,A,Y)$, where the covariates $W$ include possible confounders, mediators, and other variables. The relationships among the variables in $X$ may be thought to correspond to a DAG (or latent projection of a DAG), with the true graph partially or entirely unknown. Since distinct graphs may lead to distinct identification results and/or estimators (adjustment sets, etc.)\ for the quantity of interest, the hope is to select a submodel of the maximally agnostic supermodel where $p(x)$ is unrestricted, such that the selected restrictions are informative about strategies of identification or estimation. The paradigmatic example of ``selection'' is the causal observational epidemiologic study wherein the goal may be to estimate the average causal effect of $A$ on $Y$ and there is substantial uncertainty about which subsets of $W$, if any, are valid/efficient adjustment sets to use with standard effect estimators. There is a substantial theoretical and applied epidemiologic literature on using DAGs to identify causal effects and/or select among possible adjustment sets, but in this literature is common to assume the DAG is known precisely due to background knowledge or theory. Missing edges in these DAGs correspond to testable conditional independence constraints on $p(x)$, but DAGs constraints are rarely tested and DAGs are rarely selected on the basis of the data in epidemiologic research \citep{tennant2021use, petersen2021data}. When there is uncertainty about features of the graph (including missing edges, causal order, or impact of unmeasured confounders), this uncertainty may be reduced on the basis of the data by selecting a submodel of a more \emph{statistically} agnostic model (i.e., one that imposes fewer, if any, constraints on the distribution $p(x)$). The maximally agnostic model is represented by a complete graph, i.e., a graph with no missing edges and thus implying no conditional independence constraints at all. Submodel selection amounts to selecting some subgraph of this agnostic model. 

These two research contexts are neither exhaustive nor mutually exclusive -- studies that employ causal structure learning may not fit squarely into only one of these categories or in either of them. Indeed, local research questions within the same discipline or even within the same research project may exhibit aspects of either or both contexts. For example, uses of causal structure learning in neuroscience sometimes resemble ``discovery'' when the goal is to estimate networks of functional neural connections from among many possible structures \citep{sanchez2019estimating,dubois2020causal}, and they may resemble ``selection'' when the focus is modeling and quantifying the strengths of a small number of specific functional neural relationships, for example by mediation analysis \citep{rawls2021integrated}.  However, the rough distinction between these contexts is useful because it serves to highlight which aspects of algorithm performance are more salient in a given setting. In the context of discovery, precision (avoiding false positive causal connections) is usually more important since these are costly from the perspective of subsequent intervention development or experimental planning. Indeed, in many of the genetic applications cited above, the algorithmic procedures were designed or tuned in a way to ensure high precision, for example by using stability selection to control FDR \citep{stekhoven2012causal}. 
In contrast, in the context of selection, high recall (avoiding false exclusions, i.e., model constraints that are not supported by the data) is more important, since false independence assumptions may lead to biases in the subsequent estimation of causal effects or parameters. 

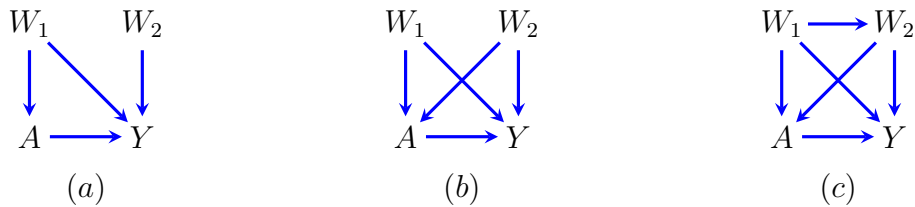
\begin{figure*}[htbp]
	\begin{center}
		\begin{tikzpicture}[>=stealth, node distance=1.5cm]
			\tikzstyle{format} = [very thick, circle, minimum size=5.0mm, inner sep=0pt]
			\begin{scope}
				\path[->, very thick]
				node[format] (w1) {$W_1$}
				node[format, right of=w1] (w2) {$W_2$}
				node[format, below of=w1] (a) {$A$}
				node[format, right of=a] (y) {$Y$}
				(w1) edge[blue] (a)
				(w1) edge[blue] (y)
				(a) edge[blue] (y)
				(w2) edge[blue] (y)
				
				node[below of=a, yshift=0.8cm, xshift=0.75cm] (l) {$(a)$};
			\end{scope}
			\begin{scope}[xshift=5cm]
				\path[->, very thick]
				node[format] (w1) {$W_1$}
				node[format, right of=w1] (w2) {$W_2$}
				node[format, below of=w1] (a) {$A$}
				node[format, right of=a] (y) {$Y$}
				(w1) edge[blue] (a)
				(w1) edge[blue] (y)
				(a) edge[blue] (y)
				(w2) edge[blue] (y)
				(w2) edge[blue] (a)
				
				node[below of=a, yshift=0.8cm, xshift=0.75cm] (l) {$(b)$};
			\end{scope}
			\begin{scope}[xshift=10cm]
				\path[->, very thick]
				node[format] (w1) {$W_1$}
				node[format, right of=w1] (w2) {$W_2$}
				node[format, below of=w1] (a) {$A$}
				node[format, right of=a] (y) {$Y$}
				(w1) edge[blue] (a)
				(w1) edge[blue] (y)
				(a) edge[blue] (y)
				(w2) edge[blue] (y)
				(w2) edge[blue] (a)
				(w1) edge[blue] (w2)
				
				node[below of=a, yshift=0.8cm, xshift=0.75cm] (l) {$(c)$};
			\end{scope}
		\end{tikzpicture}
	\end{center}
	\caption{
		(a) A DAG representing a submodel of the agnostic model in which $W_1 \independent W_2$ and $A \independent W_2$. $W_1$ alone is a valid adjustment set for the effect of $A$ on $Y$. (b) A supermodel of the DAG in (a) that does not imply $A \independent W_2$ and in which $(W_1, W_2)$ is a valid adjustment set. (c) A supermodel of both (a) and (b) that imposes no independence constraints on $p(w,a,y)$.
	}
	\label{fig:submodel}
\end{figure*}

For an illustration of this last point, consider the following hypothetical example. Let $A, Y$ denote an exposure and outcome of interest, respectively, and $W$ a set of pre-exposure covariates. Assume the target quantity is the expected outcome under an intervention that sets $A=a$ (a.k.a.\ counterfactual mean), $\E[Y(a)]$, and that there is no unmeasured confounding, i.e., we have the conditionally ignorable causal model where $Y(a) \independent A \mid W$. This model imposes no restrictions on the distribution $p(w,a,y)$. The analyst may further propose the model illustrated in Figure \ref{fig:submodel}(a), where according to the rules of sufficient adjustment, they need only adjust for $W_1$ but not $W_2$. That is, they can obtain a consistent estimate of their effect by using the g-formula, inverse probability weighting, or doubly-robust augmented inverse probability weighting entirely ignoring $W_2$, only including $W_1$ as appropriate in the outcome regression model or propensity score (or both). This model asserts that $W_1 \independent W_2$ and $A \independent W_2$, i.e., it is a submodel of the agnostic model. If this posited model is correct, then there is no issue with adjusting for only $W_1$. However, if the model is incorrect, and instead the data was generated according to a DAG that is a supermodel of Figure \ref{fig:submodel}(a), say the one displayed in Figure \ref{fig:submodel}(b), then failing to adjust for $W_2$ would lead to confounding bias in the estimation of $\E[Y(a)]$ (specifically, if $W_2$ is neither marginally nor conditionally independent of $A$). The supermodel in Figure \ref{fig:submodel}(b) would imply that both $W_1$ and $W_2$ should be adjusted for in any consistent effect estimator. The model in Figure \ref{fig:submodel}(b) is also a submodel of the agnostic conditionally ignorable model, which would correspond to a complete graph with $W_1,W_2$ adjacent: Figure \ref{fig:submodel}(c). Following the implications of any of these supermodels of the analyst's model would produce consistent estimates of the causal effect, whereas following the implications of the wrong submodel (the analyst's model or any submodel thereof) would produce bias. So, a more ``cautious" or ``conservative'' model selection procedure would prefer one of the supermodels in the absence of compelling evidence in favor of the analyst's submodel.

The upshot is that for at least some causal inference tasks, it is desirable to select models in a way that errs on the side of supermodels (fewer non-adjacencies, higher recall) rather than maximizing precision. In the context of estimating causal effects in fully-observed DAG models, selecting a supermodel of the truth may have some efficiency costs (estimators that make use of true independence constraints may be more efficient; see, e.g., \citet{rotnitzky2020efficient}), but will not lead to bias via implying incorrect adjustment sets. This remains true for settings with covariates that may also be post-exposure. This is observation may be stated in the form of a theorem. First recall that by definition, a set $W$ is an adjustment set relative to $(A,Y)$ in graph $\G$ if the ``adjustment formula'' holds with this choice of set, i.e.,
\begin{equation*}
	p(y(a)) = \begin{cases}
		p(y \mid a) & W = \emptyset \\
		\int_w p(y \mid a, w) dP(w) & \text{otherwise}
	\end{cases}
\end{equation*}
See \citet{pearl2009causality, maathuis2015generalized} and \citet{perkovic2018complete}. Then:

\begin{Thm}\label{thm:1}
	Let $X=(W,A,Y)$ be a random vector partitioned into covariates $W$, exposure $A$, and outcome $Y$, and assume $p(x)$ satisfies the Markov condition wrt (``true'') DAG $\G_0$. Let $\G_1$ denote another DAG such that $\G_0 \subseteq \G_1$. If 
	$W_1 \subseteq W$ is an adjustment set relative to $(A,Y)$ in $\G_1$, then $W_1$ is also an adjustment set relative to $(A,Y)$ in $\G_0$.
\end{Thm}

This result follows from Proposition 3 in \cite{peters2015structural}, but for completeness is also proved using only elementary graphical concepts in the Appendix. The statement implies that using a super-DAG of the true DAG to select an adjustment set will always be a valid procedure. (Though it is not the focus of this work, a similar statement could be made for identifying the effects of time-dependent treatments or ``sequential plans'' that are not identified by simple time-independent adjustment but identified via the g-formula using the graphical criteria in \citet{pearl1995probabilistic} or \citet{dawid2008identifying}.)

When there is possible unmeasured confounding, the issue is more complicated due to the fact that a causal effect identifiable in a submodel may not be identifiable in a supermodel. But in that setting too, it may be prudent to err on the side of safety: select a supermodel in which the effect of interest may not point-identified and then estimate bounds or use sensitivity analysis \citep{robins2000sensitivity, zhang2022partial, duarte2023automated}, rather than confidently assert a point estimate for a quantity that relies on identification assumptions not supported by the data.

Of course, even if recall is valued more highly than precision (or vice versa) in a given context, some balance is desirable. An easy way to obtain perfect recall while entirely sacrificing precision is to return a complete graph, which would be totally uninformative, and likewise it would be unacceptable to return an empty graph with perfect precision and dismal recall. Instead, one may hope that the structure learning procedure reflects something near the true (unknown) density of the underlying graph, with a preference for supergraphs in the context of selection.

In the following section, we propose a simple modification to classic constraint-based learning algorithms that prioritizes recall, but where the preference for denser graphs is ``tunable.''

Closely related to the above considerations is an alternative approach to quantifying the ``distance'' between an estimated graph $\widehat{\G}$ and the truth $\G_0$ proposed by \cite{peters2015structural} called the structural intervention distance (SID). Assuming a context in which $\widehat{\G}$ will be used to inform the computation of causal effects, the SID essentially counts the number of causal effect identification ``errors'' implied by $\widehat{\G}$ relative to $\G_0$ for all possible cause-effect pairs in a graph. The fundamental idea here is that the estimated structure may disagree with the true structure in ways that do not impact effect identification results, in which case $\widehat{\G}$ may be considered pragmatically adequate. If using $\widehat{\G}$ to inform identification of interventional densities for all possible cause-effect pairs in a graph yields few errors, SID will be close to zero; otherwise higher SID corresponds to a larger number of identification formula mistakes. Theorem 1 above reflects that the SID between the truth and any super-DAG of the truth is exactly zero. More recently, \citet{henckel2024adjustment} have generalized this idea and proposed a family of structural graph distances, a different distance for each possible choice of adjustment set (e.g., choosing to adjust for the parents of exposure, the ancestors of exposure, or a more complicated optimal adjustment set) -- they call these structural adjustment distances. Their proposed distances may also be applied when the estimated graph is a CPDAG. The advantage of using these distances, instead of precision or recall, for benchmarking graphical structure learning algorithms is that they are more directly tied to the downstream use of an estimated graph for causal identification. There are some drawbacks though: these distances cannot be used if the estimated graph is not a valid DAG or CPDAG (which may be an issue when using algorithms such as PC in finite samples, due to statistical errors) and these distances are not straightforward to generalize to other kinds of graph structures such as PAGs. Also, these distances look at all possible cause-effect pairs in a graph which may be less relevant if only one or a few exposures are of interest (though, precision and recall are also ``global'' measures). We revisit these structural distances in Section 5.

\section{An Equivalence Testing Approach to Improve Recovery of Edges}
In the standard approach to constaint-based graphical model selection, the null hypothesis evaluated for every choice of variables $i, j$ and conditioning set $S$ is that $X_i \independent X_j \mid X_S$. The alternative is that  $X_i \not\independent X_j \mid X_S$. This is implemented by first choosing some appropriate measure of conditional association between $X_i$ and $X_j$ given $X_S$, which generically we denote by $\theta_{ij.S}$, and testing the null hypothesis that $\theta_{ij.S} = 0$. (In practice, this is typically implemented by calculating a p-value under this null hypothesis and comparing to a threshold $\alpha$.)

We propose reversing the usual null hypothesis tests of conditional independence in constraint based algorithms, i.e., testing:
\begin{align*}
	H_0: | \theta_{ij.S} | \geq \delta \\
	H_A: | \theta_{ij.S} | < \delta 
\end{align*}
where $\delta > 0$ is a user-specified tolerance threshold. In this case, the null represents ``strong'' dependence (at least to degree $\delta$) and the alternative represents ``weak'' dependence. Starting from a complete graph, an edge would be removed only if the null hypothesis is rejected (in favor of the alternative), so the nominal Type I error being controlled is the error of a false edge removal. This reverses the testing logic of classic constraint-based algorithms and also introduces an additional parameter, the user-specified tolerance threshold $\delta$ that quantifies how ``weak'' must dependence be for it to be treated as independence.

Adopting terminology from the clinical trials literature, this may be called an equivalence testing approach. For some general results concerning tests of this form, e.g., the existence of uniformly most powerful tests, see \citet{romano2005optimal} and \citet[pp.\ 81-91]{lehmann2009testing}.
The implementation of the equivalence test will depend on what measure of conditional association is deemed appropriate for the data; see below.

The simple idea here is that when the null hypothesis of ``strong'' dependence is rejected, the algorithm accepts the alternative that whatever dependence exists is at least ``weak enough'' to be treated as zero for the purposes of model selection. When the test ``fails to reject'' the null of dependence, the procedure takes the ``cautious'' step of allowing the corresponding edge to remain, which amounts to a preference for supergraphs. A related idea was proposed in \citet{bilinski2018nothing}, where the authors use non-inferiority (one-sided equivalence) tests to detect departures from the ``parallel trends'' assumption underlying many difference-in-difference estimators. In that case, their test procedure also prefers the more ``cautious'' supermodel that does not impose the parallel trends constraint.

\subsection{Tests using partial correlation}

In settings where the distribution $p(x)$ may be assumpted to be (approximately)
multivariate Gaussian, it is typical to use Pearson's partial correlation
coefficient $\rho _{ij.S}$ as the association measure of choice. Under
the Gaussianity assumption, $\rho _{ij.S}=0$ iff
$X_{i} \protect \mathpalette{\protect \independenT }{\perp }X_{j}
| X_{S}$. In the equivalence test formulation, one would test:
\begin{align*}
	H_{0}: | \rho _{ij.S} | &\geq \delta,
	\\
	H_{A}: | \rho _{ij.S} | &< \delta .
\end{align*}
A fast estimator $\widehat{\rho}_{ij.S}$ for $\rho _{ij.S}$ is based on
inverting the empirical covariance matrix \citep{anderson2003intro}. It
is typical to base inference on a variance-stabilizing transformation of
$\widehat{\rho}_{ij.S}$ called Fisher's z-transformation:
$z(\widehat{\rho}_{ij.S}) = \frac{1}{2} \ln
\frac{1+\widehat{\rho}_{ij.S}}{1-\widehat{\rho}_{ij.S}}$. The key property
is that
$\sqrt{n-|S|-3} (z(\widehat{\rho}_{ij.S}) - z(\rho _{ij.S})) \sim N(0,1)$
approximately.

To test the above equivalence hypothesis, the procedure should reject the
null when $|z(\widehat{\rho}_{ij.S})|$ (and hence
$|\widehat{\rho}_{ij.S}|$) is small. For a two-sided level-$\alpha $ test
with fixed tolerance threshold $\delta $, the rejection rule is:
\begin{align*}
	\sqrt{n- | S | -3} \bigl( z(\widehat{\rho}_{ij.S}) - z(
	\delta ) \bigr) &< \Phi ^{-1}(\alpha )\quad \text{and}
	\\
	\sqrt{n- | S | -3} \bigl( z(\widehat{\rho}_{ij.S}) + z(
	\delta ) \bigr) &> -\Phi ^{-1}(\alpha ),
\end{align*}
where $\Phi $ is the cdf of standard normal distribution. In our implementation
discussed below, we use the (nominal)
level $\alpha =0.05$.

\subsection{Tests using the expected conditional covariance}

When the Gaussianity assumption is not warranted, it may be desirable to derive semiparametric or nonparametric tests, or, more precisely, tests that have power under some class of semiparametric or nonparametric models. No test will have power against all alternatives, and various semiparametric measures of association have been proposed. Some recent proposals have focused tests based on the expected conditional covariance, $\psi_{ij.S} = \E[\cov(X_i,X_j \mid X_S)]$ which equals zero when $X_i \independent X_j \mid X_S$ \citep{shah2020hardness}. \citet{xiang2020flexible} study a scaled version of this parameter which is takes on values in $[-1,1]$ and thus more analagous to partial correlation: $\psi^*_{ij.S} = \frac{\psi_{ij.S}}{\sqrt{\E[\sigma^2(X_i \mid X_S)] \E[\sigma^2(X_j \mid X_S)]}}$ where $\sigma^2(\cdot \mid X_S)$ denotes conditional variance given $X_S$. Estimators for the functionals $\psi_{ij.S}$ or $\psi^*_{ij.S}$ involve estimating several nuisance functionals, i.e., conditional means $\E[X_i \mid X_S]$ and $\E[X_j \mid X_S]$ (as well as conditional variances for the scaled $\psi^*_{ij.S}$). The idea behind several specific proposals \citep{shah2020hardness,xiang2020flexible} is that one may use flexible semiparametric or nonparametric estimators for these nuisance functionals, e.g., kernel regression methods, local polynomial regressions, random forests, or other prediction algorithms. Under some conditions on the convergence rates of these nuisance estimators, it is possible to construct an overall estimator $\widehat{\psi}^*_{ij.S}$ of $\psi^*_{ij.S}$ that is consistent and asymptotically normal, i.e., $\sqrt{n}(\widehat{\psi}^*_{ij.S}-\psi^*_{ij.S}) \sim N(0,\sigma_{ij.S}^2)$. The overall estimator is based on the influence function of $\psi^*_{ij.S}$ and the estimation of the asymptotic variance $\sigma_{ij.S}^2$ follows from basic semiparametric theory. 

Similarly to the case of partial correlation, we propose testing:
\begin{align*}
	H_0: | \psi^*_{ij.S} | \geq \delta \\
	H_A: | \psi^*_{ij.S} | < \delta .
\end{align*}

Using one of the aforementioned estimators, the rejection rule takes the form:
\begin{align*}
	\sqrt{n} (\widehat{\psi}^*_{ij.S}-\delta)/ \widehat{\sigma}_{ij.S}^2 &< \Phi^{-1}(\alpha) \\
	&\mbox{and}\\
	\sqrt{n} (\widehat{\psi}^*_{ij.S}+\delta)/ \widehat{\sigma}_{ij.S}^2 &> - \Phi^{-1}(\alpha)
\end{align*}

\subsection{Tests using odds ratios}

For mixed variable sets (i.e., containing continuous and binary variables) one common measure of association is the conditional odds ratio $\mbox{OR}(X_i, X_j \mid X_S)$ which equals $1$ when $X_i \independent X_j \mid X_S$. \citet{tchetgen2010doubly} and \citet{tan2019doubly} describe doubly-robust estimators for the log odds ratio $\gamma_{ij.S} = \log \mbox{OR}(X_i, X_j \mid X_S)$ in a semiparametric model, which allows for complex nonlinear relationships among the variables. Their proposed estimators are also asymptotically normal $\sqrt{n}(\widehat{\gamma}_{ij.S}-\gamma_{ij.S}) \sim N(0,\omega_{ij.S}^2)$, so the equivalence test may proceed similarly as to the previous example:

\begin{align*}
	H_0: | \gamma_{ij.S} | \geq \delta \\
	H_A: | \gamma_{ij.S} | < \delta .
\end{align*}
with rejection rule of the form:
\begin{align*}
	\sqrt{n} (\widehat{\gamma}_{ij.S}-\delta)/ \widehat{\omega}_{ij.S}^2 &< \Phi^{-1}(\alpha) \\
	&\mbox{and}\\
	\sqrt{n} (\widehat{\gamma}_{ij.S}+\delta)/ \widehat{\omega}_{ij.S}^2 &> - \Phi^{-1}(\alpha)
\end{align*}
Note that log odds ratios are not bounded in $[-1,1]$ and so the interpretation of the equivalence threshold should be modified accordingly. 

\subsection{Other independence tests}

There is a large and growing literature on semiparametric and nonparametric conditional independence testing \citep{liu2009nonparanormal,zhang2011kernel, wang2015conditional, huang2016flexible,li2020nonparametric,petersen2021testing,cai2022distribution}. Many of these proposed methods have been used in conjunction with graphical structure learning algorithms. If the test is based on a consistent and asymptotically normal estimator of some population-level parameter that measures conditional association, it may be easily transformed into an equivalence test in the same way as already described. However, some tests are based on a test statistic with known sampling distribution only under the (usual) null hypothesis of no association -- these tests cannot be straightfowardly adapted into equivalence tests unless the distribution may be characterized under alternative (dependence) hypotheses. For example, the kernel-based independence test of \citet{zhang2011kernel} is based on a test statistic with a distribution known only under the null of independence.

It is worth emphasizing that in the equivalence testing formulation of PC (or FCI, etc.), an edge is removed if some conditional association \emph{as captured by the chosen statistic or measure of association} is found to be small. Thus, missing edges will not correspond exactly to conditional independence in the fully nonparametric sense. The conditional association may be small according to some chosen measure, but larger according to another measure. This consideration arises in the traditional approach to conditional independence testing too: since no actual test will have power against all alternatives \citep{shah2020hardness}, an edge may be removed in the course of PC (or FCI, etc.) just because the chosen test statistic has insufficient power to detect a certain kind of association. 

\subsection{Choosing the equivalence threshold $\delta$}

There are several ways to choose the equivalence threshold $\delta$. One option would be to fix this choice on the basis of background knowledge about ``weak'' versus ``substantive'' conditional associations in the given context and dataset. There may be, for example, background knowledge that suggests partial correlations less than $\pm 0.10$ are not of substantive interests in a particular application, but in another setting (with moderate sample sizes or weak but important effects), partial correlations as small as $\pm 0.03$ may be important to detect. Intuitively, the choice of $\delta$ should depend on sample size, since the statistical ability to discriminate small associations from zero will grow with $\sqrt n$. Indeed, in our simulations below, we condsider choosing $\delta = C/\sqrt n$ for various choices of constant $C$. We find that for our simulation settings, choosing $C$ in the range $[1.5, 2]$ works well. We also introduce a principled \emph{data-driven} approach to selecting $\delta$ below.

\section{Empirical Performance in Finite Samples}
We evaluate the proposed approach using simulated data. We restrict our attention to the Gaussian setting so we can use tests based on partial correlation. 

\subsection{Estimating adjacencies in random graphs}

\begin{figure*}
	\includegraphics[scale=.40]{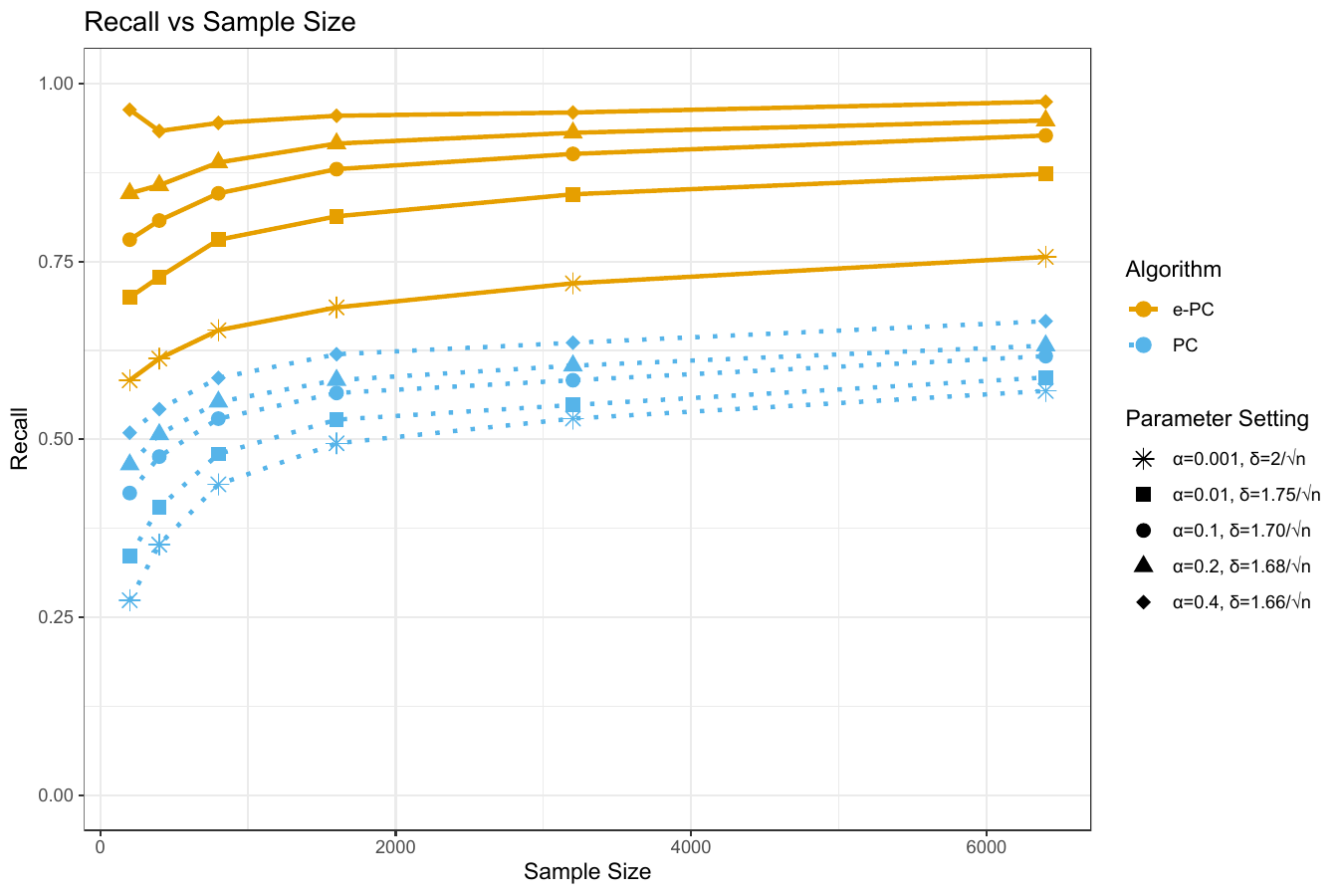}
	\includegraphics[scale=.40]{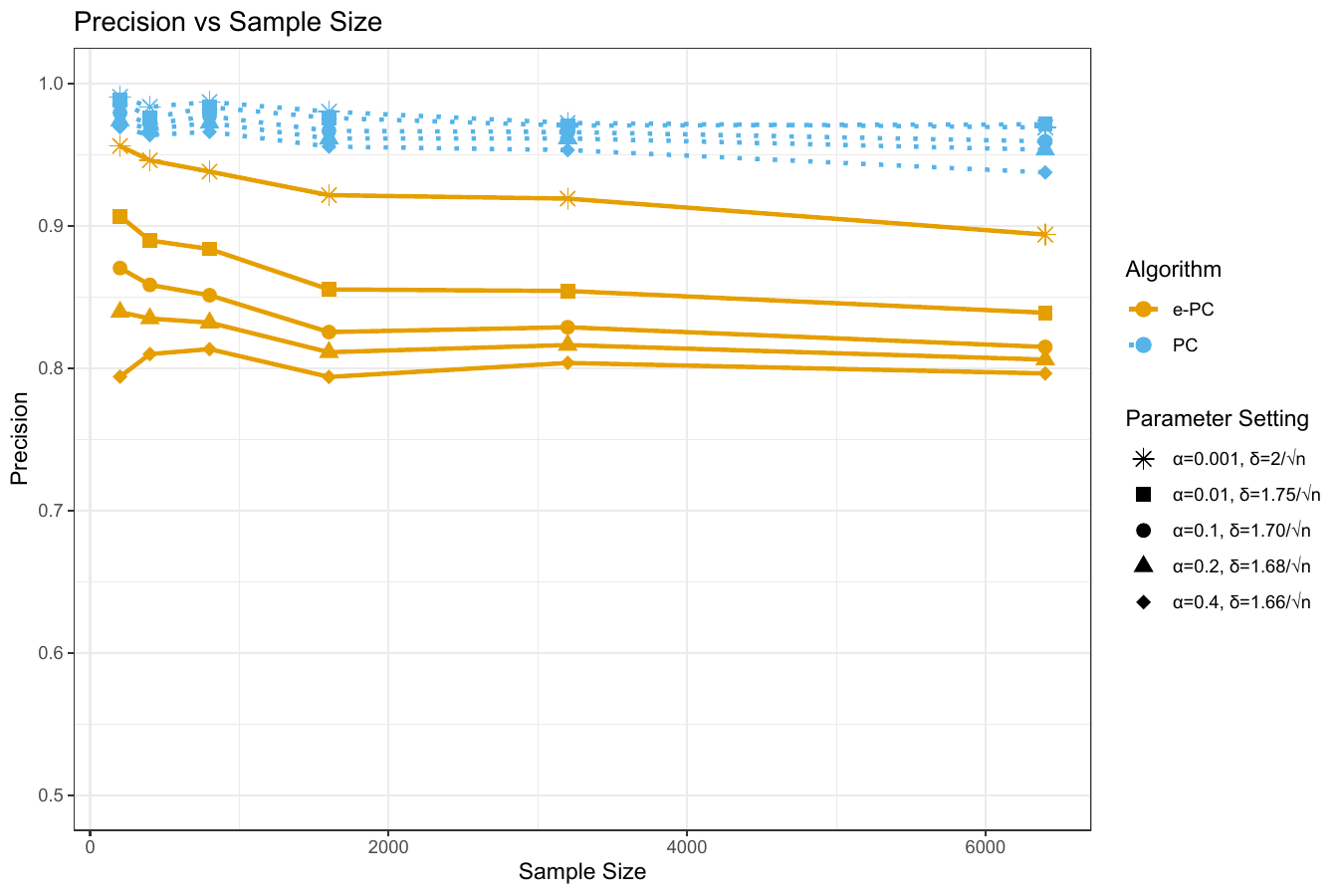}
	\caption{
		Adjacency recall and precision for the traditional and equivalence formulations of PC on simulated graphs with expected density $d=7$ and $p=10$ nodes.
	}
	\label{fig:adj}
\end{figure*}

\begin{figure*}[htbp]
	\begin{center}
		\begin{tikzpicture}[>=stealth, node distance=1.9cm]
			\tikzstyle{format} = [very thick, circle, minimum size=5.0mm, inner sep=0pt]
			\begin{scope}
				\path[->, very thick]
				node[format] (1) {1}
				node[format, right of=1] (2) {2}
				node[format, below of=1] (3) {3}
				node[format, below right of=3] (4) {4}
				node[format, below left of=4] (5) {5}
				node[format, below right of=5] (6) {6}
				node[format, below of=6] (7) {7}
				node[format, below left of=7] (8) {8}
				node[format, right of=8] (9) {9}
				node[format, below left of=9] (10) {10}

				(1) edge[-, blue] (3)
				(1) edge[-, blue] (4)
				(1) edge[-, blue, bend right=10] (8)
				(2) edge[-, blue] (4)
				(2) edge[-, blue] (6)
				(3) edge[-, blue] (4)
				(3) edge[-, blue] (6)
				(3) edge[-, blue, bend right=10] (8)
				(3) edge[-, blue] (10)
				(4) edge[-, blue] (5)
				(4) edge[-, blue, bend right=12] (7)
				(5) edge[-, blue] (7)
				(5) edge[-, blue] (8)
				(5) edge[-, blue, bend right=10] (9)
				(6) edge[-, blue] (8)
				(7) edge[-, blue] (8)
				(7) edge[-, blue] (10)
				(8) edge[-, blue] (10)
				(9) edge[-, blue] (10)
				
				node[below of=10, yshift=1.2cm, xshift=0.3cm] (l) {$(a)$};
			\end{scope}
			\begin{scope}[xshift=5cm]
				\path[->, very thick]
				node[format] (1) {1}
				node[format, right of=1] (2) {2}
				node[format, below of=1] (3) {3}
				node[format, below right of=3] (4) {4}
				node[format, below left of=4] (5) {5}
				node[format, below right of=5] (6) {6}
				node[format, below of=6] (7) {7}
				node[format, below left of=7] (8) {8}
				node[format, right of=8] (9) {9}
				node[format, below left of=9] (10) {10}
				
				(1) edge[-, blue] (3)
				(1) edge[-, blue] (4)
				(1) edge[-, blue, bend right=10] (5)
				(1) edge[-, blue] (6)
				(1) edge[-, blue, bend left=5] (7)
				(1) edge[-, blue, bend right=10] (8)
				(1) edge[-, blue, bend left=20] (9)
				(1) edge[-, blue, bend left=15] (10)
				(2) edge[-, blue] (3)
				(2) edge[-, blue] (4)
				(2) edge[-, blue] (5)
				(2) edge[-, blue] (6)
				(2) edge[-, blue, bend right=5] (8)
				(2) edge[-, blue] (9)
				(2) edge[-, blue, bend left=15] (10)
				(3) edge[-, blue] (4)
				(3) edge[-, blue] (6)
				(3) edge[-, blue] (7)
				(3) edge[-, blue, bend right=10] (8)
				(3) edge[-, blue] (10)
				(4) edge[-, blue] (5)
				(4) edge[-, blue] (6)
				(4) edge[-, blue, bend left=12] (7)
				(4) edge[-, blue] (9)
				(4) edge[-, blue] (10)
				(5) edge[-, blue] (6)
				(5) edge[-, blue] (7)
				(5) edge[-, blue] (8)
				(5) edge[-, blue, bend right=10] (9)
				(5) edge[-, blue] (10)
				(6) edge[-, blue] (7)
				(6) edge[-, blue] (8)
				(6) edge[-, blue] (10)
				(7) edge[-, blue] (8)
				(7) edge[-, blue] (9)
				(7) edge[-, blue] (10)
				(8) edge[-, blue] (10)
				(9) edge[-, blue] (10)
				
				node[below of=10, yshift=1.2cm, xshift=0.3cm] (l) {$(b)$};
			\end{scope}
			
			\begin{scope}[xshift=10cm]
				\path[->, very thick]
				node[format] (1) {1}
				node[format, right of=1] (2) {2}
				node[format, below of=1] (3) {3}
				node[format, below right of=3] (4) {4}
				node[format, below left of=4] (5) {5}
				node[format, below right of=5] (6) {6}
				node[format, below of=6] (7) {7}
				node[format, below left of=7] (8) {8}
				node[format, right of=8] (9) {9}
				node[format, below left of=9] (10) {10}
				
				(1) edge[-, blue] (3)
				(1) edge[-, blue] (4)
				(1) edge[-, blue, bend right=10] (8)
				(2) edge[-, blue] (3)
				(2) edge[-, blue] (4)
				(2) edge[-, blue] (6)
				(2) edge[-, blue, bend left=10] (7)
				(2) edge[-, blue, bend right=5] (8)
				(2) edge[-, blue, bend left=15] (10)
				(3) edge[-, blue] (4)
				(3) edge[-, blue] (6)
				(3) edge[-, blue] (7)
				(3) edge[-, blue, bend right=10] (8)
				(3) edge[-, blue] (9)
				(3) edge[-, blue] (10)
				(4) edge[-, blue] (5)
				(4) edge[-, blue, bend left=12] (7)
				(4) edge[-, blue] (9)
				(4) edge[-, blue] (10)
				(5) edge[-, blue] (6)
				(5) edge[-, blue] (7)
				(5) edge[-, blue] (8)
				(5) edge[-, blue, bend right=10] (9)
				(6) edge[-, blue] (8)
				(6) edge[-, blue] (10)
				(7) edge[-, blue] (8)
				(7) edge[-, blue] (9)
				(7) edge[-, blue] (10)
				(8) edge[-, blue] (10)
				(9) edge[-, blue] (10)
				
				node[below of=10, yshift=1.2cm, xshift=0.3cm] (l) {$(c)$};
			\end{scope}
		\end{tikzpicture}
	\end{center}
	\caption{
		Representative examples of skeletons recovered in the simulation at $n=6400$. (a) A skeleton recovered using the traditional formulation of PC at $\alpha=0.20$. (b) A skeleton recovered using the equivalence formulation e-PC at $\delta=1.66/\sqrt{n}$. (c) The true skeleton. 
	}
	\label{fig:tripdych}
\end{figure*}
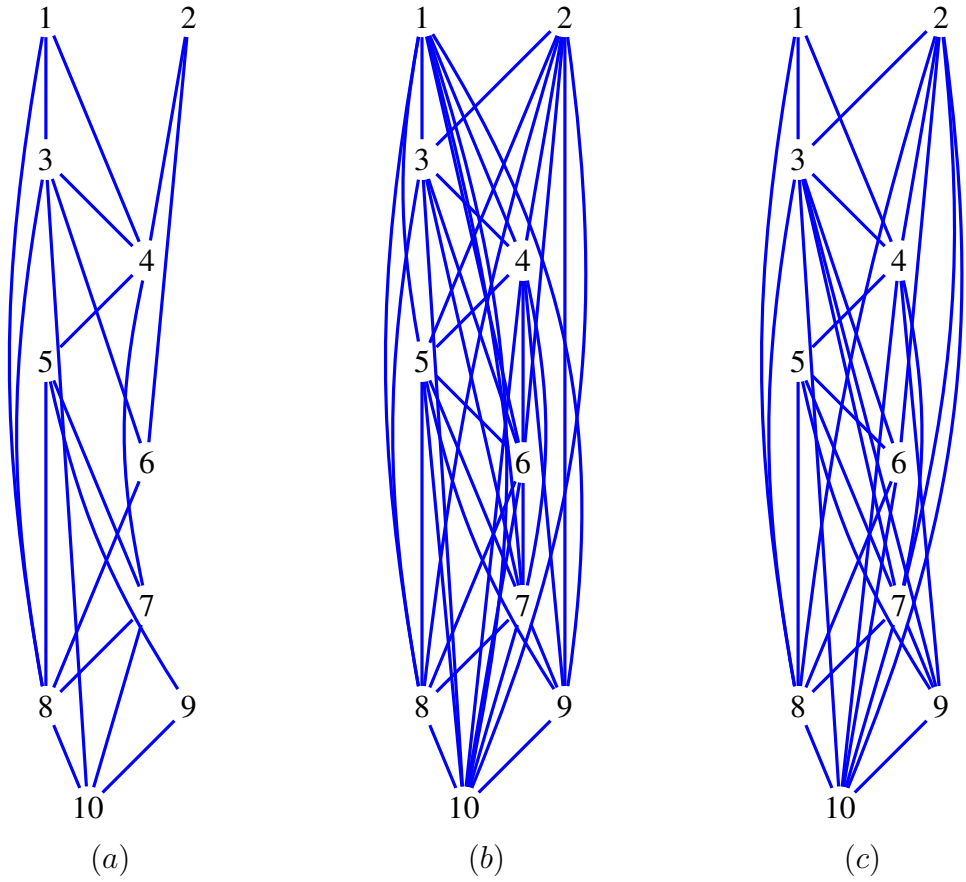

We generate random graphs from a directed Erdos-Reyni model with $p=10$ nodes with an expected density (degree) $d=7$. These are very dense graphs missing only a small number of edges ($<10$ missing out of $45$). We generated data from a linear system of equations: $X_j = \sum_{k \in \Pa_j(\G)} w_{jk} X_k + \epsilon_j$ where $\epsilon_j \sim N(0,1)$ and $w_{jk}$ are initial ``edge weights'' sampled uniformly from $\pm [0.5,1]$. In order to mitigate known simulation-dependent artifacts, namely, increasing marginal variances and correlations along the causal order \citep{reisach2024scale}, initial weights are rescaled following a normalization procedure described in \citet{mooij2020joint}, wherein each column of the weight matrix $w_{\cdot k}$ is divided by $\sqrt{||w_{\cdot k}||^2 + 1}$. Data is generated according to this procedure for a range of sample sizes $n \in \{200, 400, 800, 1600, 3400, 6800\}$ and the graph skeleton is estimated by PC. In Figure \ref{fig:adj}, we compare the performance of the traditional null hypothesis test and the proposed equivalence testing formulation (labeled ``e-PC'' in the Figure). Though e-PC has two tuning parameters that may vary, here $\alpha$ is fixed at $0.05$ while $\delta$ varies as a function of sample size. For traditional PC, only $\alpha$ varies. In both cases, we limit the maximum size of the conditioning set to $3$ to avoid tests with low power. For even large values of $\alpha$, e.g., $\alpha=0.20$ or $\alpha=0.40$, traditional PC has poor (adjacency) recall, missing up to 40\% of the true edges even at large sample sizes. e-PC is easy to tune to have excellent recall performance by choosing $\delta$ appropriately, even at small sample sizes. Of course, there is a tradeoff: since e-PC prefers dense graphs, it will produce graphs with additional edges in comparison with the truth, i.e., less precision. For these simulated graphs, a complete graph would have a mean precision of $\approx 64\%$ so the estimates from e-PC even with the smallest values of $\delta$ are not complete, but typically missing some edges. We display some representative estimated skeletons (and the truth) in Figure \ref{fig:tripdych}. The true skeleton in (c) has $30$ edges. The graph estimated by traditional PC (with $\alpha=0.20)$ is much too sparse, with only $19$ edges. The graph estimated by e-PC (with $\delta=1.66/\sqrt{n}$) has $38$ edges out of the possible $45$ (e.g., with edges missing correctly between nodes 1 and 2, 3 and 5, 6 and 9, 8 and 9, etc.). In this example the sample size was $n=6400$.

An alternative approach to estimating dense graphs would be to use the traditional formulation of PC but with p-value thresholds near $1$, e.g., $\alpha = 0.8, 0.9, 0.95,$ or higher. Indeed, in our simulations using these thresholds with dense true graphs does produce estimated graphs with average performance similar to low values of $\delta$ (not shown). However, in general these p-values corresponding to the usual null correlation test are not a reliable guide to the strength of correlations, especially at moderate sample sizes, since small correlations may correspond to non-extreme p-values. Furthermore, there is no obvious way to calibrate or select appropriately high (i.e., cautious but informative) p-value thresholds, except perhaps via a back-calculation that effectively selects $\delta$ first. 

It is noteworthy that the performance of (traditional) PC is very poor in terms of recall in this dense setting, even though the data-generating process is linear-Gaussian and there are no unmeasured confounders. This raises the concern that using traditional PC as part of a two-step ``selection $+$ inference'' process (i.e., model selection followed by identification and estimation of causal effects based on the model) is likely to be error-prone when the truth is dense. We show in the next section that e-PC is better suited to this purpose.

\subsection{Distinguishing between confounders and mediators for estimating causal effects}

Estimating an accurate skeleton or super-skeleton is important, but in some applications of causal model selection the crucial step is correctly distinguishing variables that are possible confounders, which ought to be adjusted for, from mediators, i.e., variables on the causal pathway from exposure to outcome which ought not be adjusted for. We illustrate the advantage of e-PC for this task in a simple simulation. As in the previous section, we generate random Erdos-Reyni DAGs with $10$ nodes and an expected density of $7$. We draw multivariate Gaussian data from these DAGs using the same kind of linear model as described previously and estimate graphs from each dataset using e-PC and PC. Our aim is to estimate the ATE of one node on another, namely the effect of $X_5$ on $X_{10}$ where the DAG is topologically ordered: $X_{i} < X_{i+1}$ where $i+1$ is a non-ancestor of $i$ (for $i=1,\dots,9$). So, in the true DAG $\G$ nodes $(X_1,X_2,X_3,X_4)$ are ``pre-exposure'' (i.e., possible confounders) and nodes $(X_6,X_7,X_8,X_9)$ are ``post-exposure'' (but possibly ancestors of the ``outcome'' $X_{10}$).  We simulate $200$ trials of sample size $n \in \{500, 5000\}$ and each time estimate the graph using e-PC and PC. In estimating the ATE, recovering both correct adjacencies and orientations are relevant. The accuracy of PC in recovering orientations is much improved when some time-ordering information is available \citep{petersen2021data}, so we impose partial ordering information; $(X_1,X_2) < (X_3,X_4,X_5,X_6,X_7) < (X_8,X_9) < X_{10}$. To impose this ``tiered'' time-ordering on the PC algorithm, we use the package \texttt{tpc}, implementing ``tiered'' PC \citep{andrews2021practical}.

For each selected graph, we estimate the ATE of $X_5$ on $X_{10}$ using the IDA procedure, which performs back-door adjustment based on the estimated graph \citep{maathuis2009estimating}. In the case that the estimated graph has undirected edges, i.e., represents several Markov-equivalent DAGs, IDA produces a multiset of estimated effects $\widehat{\beta} = (\widehat{\beta}^{min},...,\widehat{\beta}^{max})$, where each estimate corresponds to some DAG in the selected equivalence class. These multisets have been used to provide an interval bound or lower-bound on the true causal effect \citep{maathuis2010predicting,stekhoven2012causal}. To measure estimation error, we compare the estimated multiset of causal effects with the ``oracle'' estimate $\widehat{\beta}^{oracle}$ computed from regressing $X_{10}$ on $X_5$ and the true parents of $X_5$ in $\G$. We quantify the difference between the estimated set and oracle estimate using Int-MSE (``interval mean squared error'': an adaption of mean-squared error to interval estimates) \citep{malinsky2017estimating}. This defines a measure of interval error to be $0$ if $\widehat{\beta}^{min} < \widehat{\beta}^{oracle} < \widehat{\beta}^{max}$ and $ \min \left( (\widehat{\beta}^{oracle} - \widehat{\beta}^{min})^2 , (\widehat{\beta}^{oracle} - \widehat{\beta}^{max})^2 \right)$ otherwise. Int-MSE is the mean of these distances over the $200$ trials. Results are presented in Table \ref{table:IDA}. Effect estimates based on the output of e-PC are closer to the oracle estimate. The estimated intervals are also wider, as expected, since e-PC is more likely to return a larger equivalence class. In fact, traditional PC for the most part here estimates a unique DAG rather than a CPDAG, but often the wrong one, so the resulting estimate is unique but far from the truth. In other words, viewed as ``interval estimates,'' the output of IDA with e-PC is more ``cautious'' but more likely to contain the true (oracle) value of the ATE. The results presented are for fixed choices of $\delta = 1.68/\sqrt{n}$ and $\alpha=0.20$, though substantively similar results were obtained for other choices of tuning parameters (not shown).

\begin{table}
	\centering
	\begin{tabular}{||c c c c c||} 
		\hline
		& e-PC & e-PC & PC & PC  \\ [0.5ex]
		& Int-MSE &Width & Int-MSE &Width \\
		\hline\hline
		$n=$ 500 & 0.011 & 0.16 & 0.028  & 0.008  \\ 
		\hline
		$n=$ 5000 & 0.005 & 0.22 & 0.022  & 0.014 \\ 
		\hline
	\end{tabular} \\ [2ex]
	\caption{Int-MSE and interval width of estimated ATE of vertex $X_5$ on $X_{10}$ using IDA, across $200$ random $10$-node graphs. Here e-PC uses $\delta=1.68/\sqrt{n}$ and PC uses $\alpha=0.20$.}
	\label{table:IDA}
\end{table}

\subsection{Evaluating performance with a structural adjustment distance}

We also compare the performance of e-PC to PC using one of the structural adjustment distances introduced in \citet{henckel2024adjustment}. Specifically, we use their ``parent adjustment distance,'' which assumes that for each possible cause-effect pair in the graph, the effect will be identified by adjusting for the parents of the purported cause. As already discussed, this benchmark evaluates an estimated structure in a way that is more directly tied to its use in identifying causal effects. Using the same data and settings reported in the previous section, we find that e-PC outperforms PC in terms of parent adjustment distance. When normalized such that all distances are in $[0,1]$, the mean distance to the truth is $0.65$ for e-PC and $0.73$ for PC at $n=500$; the means are $0.60$ and $0.68$ at $n=5000$. Note that since the structural adjustment distances are not defined when the estimated graph is not a valid DAG or CPDAG, we define the distance to be $1.0$ (maximally bad) whenever this occurs. This occured more frequently for PC than e-PC in our simulations. (Improved performance of e-PC is not consistently evident in these simulations when using another distance proposed in \citet{henckel2024adjustment}, the ``Oset adjustment distance,'' which uses the variance-optimal adjustment set as the relevant identification strategy for all causal effects.)

\section{Theoretical Guarantees}
For fixed sample size $n$, denote the graph estimated by the equivalence-PC algorithm using partial correlation threshold $\delta$ by $\widehat{\G}_{n, \delta}$ (or $\widehat{\G}_{\delta}$ when the relevant sample size is clear). Consider the following assumptions:

\begin{itemize}
	\item[(A1)] The distribution $p(x)$ is multivariate Gaussian and faithful to DAG $\G_0$.
	\item[(A2)] $ \inf_{(i,j,S)} \{ | \rho_{ij.S} | :  \rho_{ij.S} \neq 0 \} \geq \delta_0$
	\item[(A3)] $ \sup_{(i,j,S)} | \rho_{ij.S} | < M \leq 1 $
\end{itemize}

\citet{kalisch2007estimating} make all of the above assumptions\footnote{\citet{kalisch2007estimating} combine (A2) and (A3) into one assumption which they call (A4). Their (A2) and (A3) are automatically satisfied in the fixed-dimension setting.} as well as additional assumptions relevant to their high-dimensional analysis (allowing dimension $p_n$ to grow with $n$). So, our result will be a closely related specialized version of their result, where we operate in a fixed-dimension context ($p < n$) and are interested in supergraph-consistency. The Gaussianity assumption in (A1) can be relaxed; non-parametric extensions of the Kalisch and Buhlmann consistency result can be found in \citet{chakraborty2022nonparametric}. We only focus on the Gaussian case for simplicity here, since the non-parametric case involves a more complicated independence test statistic than partial correlation. (A2) and (A3) require that the partial correlations are bounded from above and below. In contrast to the result in \citet{kalisch2007estimating}, the user of e-PC aims to choose a minimum partial correlation threshold $\delta \leq \delta_0$ (in \citeauthor{kalisch2007estimating}'s result this is a free and unknown parameter). The upper bound in (A3) is needed so that the partial correlation coefficient and its z-transform are well-defined.  

\begin{Thm}\label{thm:2}
	Assume (A1), (A2), and (A3). Let $\delta \leq \delta_0$ and $\widehat{\G}_{n, \delta}$ denote the graph estimated by equivalence-PC. There exists $\alpha_n \to 0$ such that $\mathbb{P}( \sk(\G_0) \subseteq \sk(\widehat{\G}_{n,\delta})) \to 1$ as $n \to \infty$.
\end{Thm}

This supergraph-consistency result requires specifying an appropriate threshold for the minimum conditional association (partial correlation) size. Next, more interestingly, we describe a ``monotonicity'' property that is useful for choosing $\delta$ in practice.

\begin{Thm}\label{thm:3}
	Assume (A1), (A2), and (A3). Let $\delta_0$ denote the true minimal partial correlation size in (A2). 
	For two graphs $\widehat{\G}_{n, \delta}$ and $\widehat{\G}_{n, \delta'}$ estimated from the same sample with $\delta < \delta' \leq \delta_0$, using the sequence $\alpha_n \to 0$ from Theorem 2, $\mathbb{P}( \sk(\widehat{\G}_{n,\delta'}) \subseteq \sk(\widehat{\G}_{n,\delta})) \to 1$ as $n \to \infty$.
\end{Thm}

This monotonicity property, i.e., that \emph{so long as both considered thresholds are $\leq \delta_0$}, increasing $\delta$ to produce a more sparse graph estimate results in a subgraph, can be useful for choosing $\delta$ in practice when $\delta_0$ is unknown. Importantly, it would be difficult to even formulate a similar property for the classical formulation of PC with p-value thresholds, since there is no corresponding concept of a ``true p-value threshold.'' We propose the following procedure for selecting a choice of $\delta$ in practice:
\begin{itemize}
	\item[1.] Specify an increasing sequence (grid) of $\delta$ values: $\delta_{\min}, \cdots , \delta_{\max}$
	\item[2.] Beginning with $\delta_{\min}$, for adjacent values $\delta < \delta'$ in the sequence, estimate $\mathbb{P}( \sk(\widehat{\G}_{n,\delta'}) \subseteq \sk(\widehat{\G}_{n,\delta}))$ by subsampling the data.
	\item[3.] If this probability is no longer ``high'' ($>90$\%), then stop and select $\delta$.
\end{itemize}
Note that increasing $\delta, \delta'$ past the true $\delta_0$ will lead to increasingly sparse graph estimates, but there will be no guarantee in this regime that the edges removed correspond to true independences (or small associations), i.e., the edges may be falsely removed. The role of the requirement that $\delta < \delta' \leq \delta_0$ is that Type I error (false edge removal) is appropriately controlled ($n\to \infty$), as is detailed in the proofs of Theorem 2 and 3 in the Appendix.

Admittedly, to implement the proposed selection procedure the user must specify a threshold for what counts as ``high probability'' (we use $90$\%), but this is arguably interpretable. We implement step 2 by repeatedly subsampling the data (e.g., drawing a sample of size $0.9 n$ without replacement), estimating $\widehat{\G}_{\delta'}, \widehat{\G}_{\delta}$ on each subsample, and simply counting the frequency with which the subgraph inclusion relationship holds as a fraction of the number of repetitions. This constitutes a data-driven approach to choosing $\delta$. It is closely related to other proposed approaches to tuning-parameter selection based on subsampling or ``stability'' considerations, for example the StARS proposal of \citet{liu2010stability}. StARS selects a tuning (regularization) parameter for estimating undirected graphical models based on variability in the number of estimated edges across choices of the parameter. However, StARS was proposed for a (high-dimensional) setting in which sparsity could be assumed; in the context of our simulated dense graphs, an adaptation of this approach was found to perform quite poorly, selecting graphs that are much too sparse. Similarly, the ``stability selection'' approach of \citet{meinshausen2010stability}, which also uses subsampling, aims to control FDR (false edge inclusions) and produces sparse graphs. In contrast, the proposed approach selected graphs with a density closely matching the density of the true graph. Graphs and data of sample size $n=500$ were simulated according to the same procedure as described in Section 5. Values of $\delta$ were considered in a grid of $10$ equally-spaced values between $\delta_{\min} = 1.5/\sqrt{n} \approx 0.067$ and $\delta_{\max} = 2.5/\sqrt{n} \approx 0.11$. Over $50$ repetitions, the proposed procedure selected a $\delta \approx 0.077$ each time, which produced graphs with an average of $31.42$ edges (while the true graphs had an average of $35$) for an average recall of $\approx 79\%$. With a larger sample size of $n=5000$ and a $\delta$-grid defined the same way, the same procedure selected a $\delta$ between $\approx 0.024$ and $\approx 0.029$, which produced graphs with an average of $36.32$ edges and an average recall of $\approx 86\%$.

We conclude this section by emphasizing that these theoretical properties do not depend in any essential way on the Gaussianity assumption or using partial correlation as the measure of association; existing nonparametric consistency results for the PC algorithm, e.g., \citet{chakraborty2022nonparametric}, follow the same fundamental proof strategies (showing that the probability of errors in PC vanish as $n \to \infty$) but use more complicated test statistics for judging association. So, statements analogous to Theorems 2 and 3 would apply in those situations.

\section{Estimating effects of chemical exposures in NHANES data}

To illustrate the utility of the proposed approach, we apply the equivalence formulation of PC to data from the National Health And Nutrition Examination Survey (NHANES) in order to first select a causal model, and then estimate the effect of Perfluoroalkyl substance (PFAS) exposure on glycemic control in adults over 65 years old. Our choice of target estimand and relevant variables is informed by a recent study \citep{brosset2023exposure}. In that work, the authors investigated the association of serum concentrations of several different PFAS compounds on glycemic control in adults over 65 with known type 2 diabetes mellitus, using NHANES data from 1999-2018. Of particular interest, the authors report some uncertainty in what are valid adjustment sets for estimating the effect of interest. In their statistical analysis, they adjust for a list of selected covariates, but note that some covarites, for example related to kidney function and body mass index, ``might represent intermediate variables in the pathway from PFAS levels to poor glycemic control.'' That is, these covariates (which are known to be strongly associated with PFAS) may be possible mediators affected by exposure rather than confounders, and if so they ought not be included as adjustment variables. In their primary analysis \citeauthor{brosset2023exposure} assume all the selected covariates are confounders and adjust for them. They posit a DAG to justify this, included in their supplementary materials. The authors address their uncertainty about causal relationships among covariates by presenting several sensitivity analyses, some of which disagree substantively with each other. Our analysis is inspired by this study and we chose our sample and covariates to approximate what was reported therein. However, ours is not a direct replication because we were not able to reconstruct their exact analytical sample and we include all adults $\geq 65$, not only those with known diabetes.\footnote{Unfortunately, some of the data cleaning steps described in the paper were not sufficiently detailed to reconstruct their sample. The authors did not respond to communication.}

We focus on one PFAS compound of interest: perfluorononanoic acid (PFNA). The outcome is average level of blood glucose, HbA1C as captured in NHANES. (\citeauthor{brosset2023exposure} binarize HbA1C levels at 7\% or 8\% to define ``poor glycemic control.'' We use the continuous level.) Covariates include age at interview, sex, body mass index (BMI), blood pressure (for which we use a summary called mean blood pressure, a weighted combination of systolic and diastolic), a measure of chronic kidney disease (for which we use albumin-to-creatinine ratio in mg/g, Al2Cr), a measure of cigarette smoking (based on serum cotinine level, in ng/mL), marital status, diet (for which we use energy intake in kCal), poverty (poverty-to-income ratio), and survey wave. These covariates nearly match (but not exactly) the covariates used in \citet{brosset2023exposure}.\footnote{A few covariates described in that paper were not disambiguated from similar quantities in the NHANES data. We do not include insulin treatment since, according to the hypothesized model posited by \citeauthor{brosset2023exposure}, it is a mediator. We also do not include a second measure of kidney health, estimated glomerular filtration rate. We chose to use continuous versions of the relevant quantities wherever possible.} We extracted data on these covariates on all adults in NHANES 1999-2018 and limited our analysis to participants for whom PFNA was measured. The final sample size was $n=3255$. PFNA level, albumin-to-creatinine ratio, and HbA1c were log-transformed. No survey weights were used in this analysis.

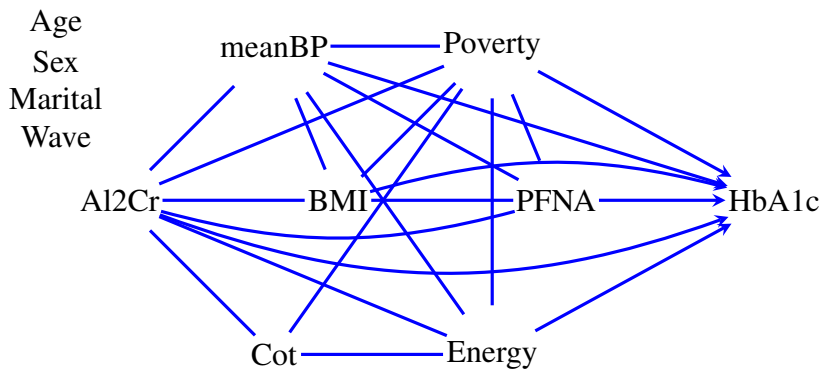
\begin{figure*}[htbp]
	\begin{center}
		\begin{tikzpicture}[>=stealth, node distance=2.9cm]
			\tikzstyle{format} = [very thick, circle, minimum size=5.0mm, inner sep=0pt]
			\begin{scope}
				\path[->, very thick]
				node[format] (al2cr) {Al2Cr}
				node[format, above right of = al2cr] (meanbp) {meanBP}
				node[format, right of = al2cr] (bmi) {BMI}
				node[format, below right of = al2cr] (cot) {Cot}
				node[format, below right of = bmi] (energy) {Energy}
				node[format, above right of = bmi] (poverty) {Poverty}
				node[format, right of = bmi] (pfna) {PFNA}
				node[format, right of = pfna] (hba1c) {HbA1c}
				
				node[left of = meanbp, yshift = 0.3cm] (age) {Age}
				node[below of = age, node distance = 0.5cm] (sex) {Sex}
				node[below of = sex, node distance = 0.5cm] (marital) {Marital}
				node[below of = marital, node distance = 0.5cm] (wave) {Wave}
				
				(al2cr) edge[-, blue] (meanbp)
				(al2cr) edge[-, blue] (bmi)
				(al2cr) edge[-, blue] (cot)
				(al2cr) edge[-, blue] (energy)
				(al2cr) edge[-, blue] (poverty)
				(al2cr) edge[-, blue, bend right = 15] (pfna)
				(al2cr) edge[blue, bend right = 20] (hba1c)
				(meanbp) edge[-, blue] (bmi)
				(meanbp) edge[-, blue] (energy)
				(meanbp) edge[-, blue] (poverty)
				(meanbp) edge[-, blue] (pfna)
				(meanbp) edge[blue] (hba1c)
				(bmi) edge[-, blue] (poverty)
				(bmi) edge[-, blue] (pfna)
				(bmi) edge[blue, bend left = 15] (hba1c)
				(cot) edge[-, blue] (energy)
				(cot) edge[-, blue] (poverty)
				(energy) edge[-, blue] (poverty)
				(energy) edge[blue] (hba1c)
				(poverty) edge[-, blue] (pfna)
				(poverty) edge[blue] (hba1c)
				(pfna) edge[blue] (hba1c)
				;
			\end{scope}
		\end{tikzpicture}
	\end{center}
	\caption{
		A graph recovered by the PC algorithm at $\delta = 2.0/\sqrt{n}$ from NHANES data described in the text. Edges from the four variables on the top-left (Age, Sex, Marital Status, and Survey Wave) are not shown to reduce clutter in the figure, but these are assumed to be parents of all other variables in the analysis.
	}
	\label{fig:PFASgraph}
\end{figure*}

Using these data, we estimated a CPDAG using e-PC, the equivalence formulation of the PC algorithm. The conditional independence judgements used partial correlation as the test statistic. In order to incorporate the discrete variables (age, sex, marital status, survey wave), all independence tests were performed on residuals from regressions of each remaining continuous variable on this set; this is equivalent to (linearly) adjusting for age, sex, marital status, and survey wave in every independence test, which is consistent with these variables being considered ``baseline'' parents of all other variables. Thus, the number of vertices considered in the course of e-PC was 8 (rather than 12), with the 4 discrete variables automatically considered parents of all other variables in the resulting graph. Since partial correlations are standardly computed from the full pairwise correlation matrix (a sufficient statistic), we used pairwise complete observations to compute the initial correlation matrix and made no further adjustment for missing data in the subsequent independence tests. (Alternative approaches to handling missing data in structure learning, e.g., test-wise deletion or multiple imputation, were not applied here but could straightforwardly be combined with e-PC, c.f.\ \citet{witte2022multiple}.) That the outcome HbA1c cannot be a causal ancestor, only possibly a descendent, of any of the other variables was imposed as (``tiered'') background knowledge on the PC algorithm. PC was run using a nominal level $\alpha=0.05$ for all equivalence tests, a maximum conditioning set size of 5, and the ``conservative'' orientation procedure \citep{ramsey2006adjacency}. The results with $\delta = 2.0/\sqrt{n} \approx 0.035$ are shown in Figure \ref{fig:PFASgraph}. At this choice of $\delta$, 6 out of 28 possible edges were removed (mostly potential edges involving cotanine levels/smoking or energy intake) and no unforced edges were oriented by the algorithm (orientations into HbA1c were imposed by the background knowledge). This means that at this choice of $\delta$ the algorithm could not by itself determine which of the covariates of interest (i.e., Al2Cr, BMI, blood pressure, energy intake, poverty) were confounders vs.\ mediators of the PFNA to HbA1c relationship. As such, the range of causal effect estimates described below cover all possible arrangements of confounders/mediators consistent with the estimated skeleton. This represents a more comprehensive range of model specifications than the sensitivity analyses presented in \citet{brosset2023exposure}. At slightly higher values of $\delta$, e.g., $\delta = 2.2/\sqrt{n}$ or $2.4/\sqrt{n}$, some additional edges are removed but the relationships among PFNA and the key covariates of interest remain mostly without determinate direction. At $\delta = 2.5/\sqrt{n} \approx 0.044$, additional edges are removed (12 total) that are sufficient to trigger PC orientation rules and orient a few directed edges, including notably BMI $\rightarrow$ PFNA and mean blood pressure $\rightarrow$ PFNA, so these would be fixed as valid adjustment variables, not mediators, at this choice of $\delta$.

Using the more dense (``cautious'') graph estimated at $\delta = 2.0/\sqrt{n}$ displayed in Figure \ref{fig:PFASgraph}, we were able to determine all possible parents of PFNA consistent with the estimated skeleton, and adjust for these parents in a series of linear mixed model regressions to estimate the effect of (log-transformed) PFNA on HbA1c in the context of a linear model. The baseline covariates age, sex, and marital status were included as adjustment variables in all specifications, and survey wave was incorporated as a random effect. The effect estimates ranged from $\beta = 0.0022$ to $\beta = 0.010$ with 16 distinct estimated values consistent with the graph. This represents a substantively wide range of estimates with a scaling factor of $\approx 4.5$ separating the minimum and maximum estimates. The accompanying nominal confidence intervals also varied, with the lower confidence limit ranging from $-0.0046$ to $0.0031$ and upper confidence limit ranging from $0.0090$ to $0.017$. The analysis was repeated with the more sparse estimated graph at $\delta = 2.5/\sqrt{n} \approx 0.044$ (not shown), and effect estimates ranged from $\beta = 0.0027$ to $\beta = 0.0089$, so conclusions (and accompanying nominal confidence intervals) are substantively similar to the primary analysis. Note that we use the same data here to perform structure learning and to estimate effects, putting aside post-selection issues. Approaches to valid post-selection inference in this ``two-step'' context are a topic of ongoing research \citep{strieder2023confidence,gradu2025valid,chang2026post}.

This analysis demonstrates how to ``cautiously'' select a causal model from observational data in order to estimate a range of causal effects. In this case, the data did not support strong conclusions about whether, for example, kidney disease (as measured by Al2Cr) acts as a confounder or mediator of the effect of interest. Several partial correlations were found to be small enough to warrant removing edges from the complete graph, but these edges were not particularly informative about variables other than smoking (serum cotanine levels). It is noteworthy that if we apply the traditional formulation of PC to the same data, with tests of the null hypothesis of zero partial correlation at the $\alpha=0.05$ level, the result is an implausibly sparse graph that removes 20 (out of 28) edges and in which PFNA is d-separated from the outcome. The result in this case would strain credulity. Even using a higher level $\alpha=.20$, the graph estimated by traditional PC is too sparse. Using the equivalence testing formulation, the procedure is sensitive to the small but potentially substantive correlation values observed in this data, and the graph recovered reflects a much smaller set of independence constraints. 

\section{Unmeasured confounding, the FCI algorithm, and other constraint-based algorithms}

The presentation so far has focused on selecting an equivalence class of DAGs, represented by a CPDAG. In the context of submodel selection, it may be more appropriate to allow for the possibility of arbitrary unmeasured confounders influencing variables in the graph. As mentioned in Section 3.1., the FCI algorithm tests conditional independence constraints to remove edges in the same way as PC but performs a potentially greater number of tests and has many more orientation possibilities to consider. The resulting equivalence class is represented by a PAG. For completeness, pseudocode for FCI is reproduced as Algorithm 8.1. The algorithm refers to a set of vertices defined for a partially oriented graph called ``possible-d-sep,'' abbreviated $\pds(i,j,\G)$. Let $k \in \pds(i,j,\G)$ if and only if $k \neq i$, $k \neq j$, and there is a path $\pi$ between $i$ and $k$ in $\G$ such that for every subpath $\langle m,l,h \rangle$ of $\pi$ either $l$ is a collider on the subpath in $\G$ or $\langle m,l,h \rangle$ is a triangle in $\G$. A \emph{triangle} is a triple $\langle m,l,h \rangle$ where each pair of vertices is adjacent. The importance of this set is that it can be proved that any possible m-separating set for nodes $i,j$ in a MAG is a subset of $\pds(i,j,\G)$. There are other valid possible definitions of $\pds(i,j,\G)$ that affect only the computational complexity of FCI, see \citet{colombo2012learning}. In the pseudocode, we use edges with an asterisk ($\ast$) to denote an arbitrary possible edge mark, e.g., $\rightarrowstar$ denotes $\rightarrow$, $\circlearrow$, or $\leftrightarrow$.
\begin{figure*}
	\begin{pseudocode}[ruled]{FCI}{\textsc{Test}, \alpha}
		\mbox{\textbf{Input:} Samples of the vector $X = (X_1,...,X_p)$} \\
		\mbox{\textbf{Output:} PAG $\mathcal{P}$} \\
		{1.} \hspace{1.5 mm} \mbox{Form the complete graph $\mathcal{P}$ on vertex set $V=\{1,...,p\}$ with $\circlecircle$ edges.} \\
		{2.} \hspace{1.5 mm} s \leftarrow 0 \\
		{3.} \hspace{1.5 mm} \mbox{\bf{repeat}}\\
		{4.} \hspace{1.5 mm} \hspace{5 mm}\FORALL \mbox{pairs of adjacent vertices $(i,j)$ s.t. $|\Adj_i(\mathcal{P})\setminus \{j\}| \geq s$}\\ 
		\hspace{9.5 mm} \mbox{and subsets $S \subseteq \Adj_i(\mathcal{P})\setminus \{j\}$ s.t. $|{S}|=s$}\\
		{5.} \hspace{1.5 mm} \hspace{10 mm} \IF \mbox{$X_i \independent X_j | X_{S}$ according to (\textsc{Test}, $\alpha$)}\\ \hspace{15 mm} \mbox{\textbf{then}} \BEGIN \mbox{Delete edge $i$ $\circlecircle$ $j$ from $\mathcal{P}.$}\\
		\mbox{Let $\mbox{sepset}(i,j) = \mbox{sepset}(j,i) = {S}$.} \END \\
		{6.} \hspace{1.5 mm} \hspace{5 mm} \mbox{\bf{end}}\\
		{7.} \hspace{1.5 mm} \hspace{5 mm} s \leftarrow s+1\\
		{8.} \hspace{1.5 mm} \mbox{\textbf{until} for each pair of adjacent vertices $(i,j)$, $|\Adj_i(\mathcal{P})\setminus \{j\}|<s$.}\\
		{9.} \hspace{1.5 mm} \FORALL \mbox{triples $(i,k,j)$ s.t. $i \in \Adj_k(\mathcal{P})$ and $j \in \Adj_k(\mathcal{P})$}\\ 
		\hspace{5 mm} \mbox{but $i \not\in \Adj_j(\mathcal{P})$, orient $i \rightarrowstar k \leftarrowstar j$ iff $k \not\in \mbox{sepset}(i,j)$.}\\
		{10.} \hspace{1.5 mm} \FORALL \mbox{pairs $(i,j)$ adjacent in $\mathcal{P}$ \textbf{if} $\exists {S}$ s.t.}\\ 
		\hspace{7.5 mm} \mbox{${S} \in \pds(i,j,\mathcal{P})$ or ${S} \in \pds(j,i,\mathcal{P})$ and $X_i \independent X_j | X_{S}$ according to (\textsc{Test}, $\alpha$)}\\
		\hspace{15 mm} \mbox{\textbf{then}} \BEGIN \mbox{Delete edge $i \starstar j$ from $\mathcal{P}.$}\\
		\mbox{Let $\mbox{sepset}(i,j) = \mbox{sepset}(j,i) = {S}$.} \END \\
		{11.} \hspace{1.5 mm} \mbox{Reorient all edges as $\circlecircle$ and \textbf{repeat} step 9.}\\
		{12.} \hspace{1.5 mm} \mbox{Exhaustively apply orientation rules (R1-R10) in Zhang (2008b) to orient}\\ \hspace{7.5 mm} \mbox{remaining endpoints.}\\
		{13.} \hspace{1.5 mm} \mbox{\textbf{return} $\mathcal{P}$}.
	\end{pseudocode}
\end{figure*}
Just as the PC algorithm, FCI uses conditional independence judgements to remove edges and orient colliders. It may perform additional tests as compared with PC and has additional, more complicated orientation rules that we omit here, referring to \citet{zhang2008completeness}. (The first four rules in \cite{zhang2008completeness} are the same as those that appear in PC, described above. Zhang's rules are sound and complete in the absence of any additional background knowledge, but the rules are not necessarily complete if additional background knowledge is imposed; see results in \citet{andrews2020completeness, venkateswaran2024towards}, and \citet{bang2025constraint}.) So, the same equivalence testing formulation of tests can straightforwardly be used with the FCI algorithm. Since the first step of FCI is identical to the adjacency search step of PC, the performance with respect to determining the presence and absence of edges is virtually the same in our simulations: using FCI with equivalence tests in place of PC on the data in Section 5.1. produces indistinguishable adjacency precision and recall results (omitted). The consequence of erring on the side of ``caution'' is that FCI with equivalence tests may recover a much more dense and less informative set of latent variable models as compared with the traditional FCI, i.e., is likely to estimate a dense PAG. In broad outline, the same basic procedure for estimating causal effects based on an PAG may be pursued: enumerate graphs in the equivalence class (or rather, enumerate valid adjustment sets) and estimate the causal effect of interest with each possible adjustment set \citep{maathuis2015generalized,malinsky2017estimating}. An important difference here is that in some or all members of the equivalence class it may be that the effect of interest is not identified. This would occur if unmeasured confounding between the exposure and outcome cannot be ruled out by independence constraints on the observed data distribution. Or, put another way, allowing for a space of models that does not \emph{a priori} rule out unmeasured confounding, the causal effect of interest will only be identified in the selected PAG if the observed pattern of conditional independence relationships is \emph{incompatible} with unmeasured confounding among the exposure and outcome. (Unmeasured confounding need not be ruled out everywhere in the graph, e.g., among all of the covariates, in order for the ATE of interest to be identifiable.) The circumstances under which independence constraints render causal effects identifiable has been described in detail elsewhere \citep{sgs2000,zhang2008causal,entner2013data}, but the key point is that observed conditional independence constraints may imply ``exclusion restrictions,'' i.e., absences of causal pathways that in conjunction (and under a faithfulness assumption) render the existence of unmeasured confounding empirically testable. The classic example in the graphical models literature is the so-called ``Y-structure'' \citep{mani2006theoretical}.
Whether an effect will be identifiable in this context then depends on a set of conditional independence judgements, which may be evaluated using the equivalence testing formulation. If the data does not support removing a sufficient set of edges to render the target effect identifiable, then the supermodel selected will be reasonably called ``cautious'' since no strong causal assertions will be supported. Here a supermodel of the truth may be uninformative about the causal question of interest, but not incorrect. 

With this in mind, the approach outlined in the previous section may applied while allowing for unmeasured confounding. We analyzed the same NHANES data using FCI (using all the same steps as above at $\delta=2.0/\sqrt{n}$): the algorithm removed the same edges as with PC and likewise oriented nothing except what was imposed by background knowledge.

Finally, we mention briefly that the equivalence testing formulation may be used in conjunction with other constraint-based or hybrid structure learning algorithms. For example, permutation-based algorithms for learning (equivalence classes of) DAGs and MAGs have been shown to perform well in practice and enjoy statistical guarantees given assumptions weaker than faithfulness \citep{solus2021consistency}. These algorithms are ``hybrid'' in the sense that they combine conditional independence tests with score comparisons. The conditional independence tests are in practice also implemented with zero association (or zero partial correlation) acting as the null, but these could be replaced with equivalence tests. Other algorithmic approaches that use conditional independence judgments, including procedures using SAT-solvers to combine independence judgments \citep{hyttinen2013discovering,zhalama2017sat}, could be similarly modified. Thus, the approach outlined here may be combined with a large class of algorithmic procedures for estimating graphs. Score-based structure learning algorithms constitute an alternative approach to learning graphical models, and enjoy desirable properties \citep{chickering2002learning, nandy2018high}. In the linear-Gaussian setting, it is sometimes possible to perform exact search to find a globally optimal (best scoring) graph, though this is typically limited to sparse structures and does not scale well to a large number of variables \citep{silander2006simple,ng2021reliable,rantanen2021maximal}. Recent advances have led to more scalable greedy algorithms that may perform better than constraint-based algorithms in terms of recall, also in the linear-Gaussian setting \citep{lam2022greedy,andrews2023fast}. It is not so straightforward how to modify score-based algorithms to prefer supergraphs of the truth in a way that enjoys similar theoretical guarantees. This may be a fruitful direction for future research.

\section{Conclusion}

Causal structure learning may serve distinct purposes across different study contexts. Approaches to improving the performance of structure learning algorithms have largely focused on what we call the context of causal discovery, where the true causal relationships may be plausibly assumed to be sparse and including false causal edges in the estimated output is associated with some high cost. In the contex of submodel selection, independence constraints and missing edges lead to possibly distinct identification results (identification strategies, adjustment sets, estimands) and so the relevant risk lies in falsely removing an edge from the more ``agnostic'' model. A procedure that is ``cautious'' in this context would prefer dense graphs and control the error of false edge removal. To this end, we propose a simple inversion of the usual null hypothesis testing approach embedded in classical constraint-based structure learning algorithms. Though the modification to existing algorithms is minor, this leads to importantly different behavior of PC and raises relevant statistical considerations, such as the choice of the equivalence threshold, which may be informed by substantive domain-specific knowledge or chosen in a data-driven way. Using the equivalence test formulation for removing edges directly affects the estimated graph adjacencies (skeleton), but also may affect recovered edge directions since in PC and FCI these are driven by the detection of unshielded colliders; removing fewer edges in the first stage of the procedure will typically lead to fewer determinate edge directions, which will broadly lead to more conservative or uncertain causal determinations, but also fewer incorrect conclusions. More broadly, the perspective outlined shifts the methodological focus away from precisely estimating the ``true'' graphical structure to an emphasis on recovering a graph that is well-suited to a downstream scientific purpose.

The results presented here suggest that, when the ultimate goal is to use the output of structure learning to select adjustment variables and estimate causal effects, the equivalence testing formulation should more reliable and more conservative. There are so far few works that systematically evaluate the reliability of this ``two-step'' discovery $+$ inference pipeline in the context of submodel selection: influential earlier work such as related to the IDA procedure \citep{maathuis2009estimating,maathuis2010predicting} has mostly focused on the context of discovery where metrics such as false discovery rates are of more interest than something that quantifies the bias of specific estimated causal effects. Some recent work has evaluated the reliability, in terms of bias, of using adjustment sets informed by graphical algorithms in a context with a known (null) causal effect of interest \citep{gururaghavendran2025can}. They find mixed results depending on what algorithmic and analytical decisions are made. More research in this vein could be informative especially when choosing among different possible algorithms and analytical pipelines. The equivalence testing formulation of constraint-based algorithms ought to be evaluated alongside other algorithms in these kinds of studies. 

The proposed subgraph ``stability'' procedure for data-driven selection of the equivalence threshold raises a general question: how best to select hyperparameters in causal structure learning when the downstream goal is to estimate causal effects. The procedure outlined is one sensible approach in that it preferentially selects a dense graph that is likely to be a super-graph of the truth and thus lead to cautious inference. It is not however known to be ``optimal'' in any sense and not specifically targeted to any specific causal effect if one such effect is the focus of inference. There is also no reason why the equivalence threshold must be the same for all variable pairs or hypotheses tested: it may be that $\delta$ is actually varies or is adaptively chosen for different independence hypotheses. It seems that more research is needed on a practical criteria for selecting tuning parameters in a way that is ``local'' and specifically tailored to the downstream inference goal.

One possible limitation of the proposed approach is its dependence on the choice of conditional association measure, e.g., partial correlation, some conditional covariance measure, or other statistic that may form the basis of independence tests. As already mentioned, the traditional approach to independence testing in causal structure learning faces this same challenge, but this manifests as an implicit choice of alternative hypotheses against which the chosen test has adequate power. Users have leeway to match their choice of independence test or association measure to the substantive problem at hand and calibrate what counts as ``small'' on that specific measure in the relevant scientific domain, which may be challenging but also lead to more transparent and credible analyses.

	\section*{Appendix}
	\begin{proof}[Proof of Theorem \ref{thm:1}]
		A path from $A$ to $Y$ is called proper if only its first node is in $A$. A possibly causal path from $A$ to $Y$ is a path from $A$ to $Y$
		that does not contain an arrowhead pointing in the direction of $A$. A path from $A$ to $Y$
		that is not possibly causal is called a non-causal path from $A$ to $Y$. Let $\mbox{Forb}(A, Y, \G)$ denote the set of all descendants in $\G$ of any node not in $A$ which lies on a proper directed path from $A$ to $Y$ in $\G$. By Definition 55 and Theorem 56 in \citet{perkovic2018complete} (see also: \citet{shpitser2010validity,van2019separators}), $W_1$ is an adjustment set for $(A,Y)$ in $\G_1$ if and only if the following two conditions hold: (i) $W_1 \cap \mbox{Forb}(A, Y, \G_1) = \emptyset$ and (ii) all proper non-causal paths from $A$ to $Y$ in $\G$ are blocked by $W_1$. Since $\G_0$ has a subset of the edges in $\G_1$, $\mbox{Forb}(A, Y, \G_0) \subseteq \mbox{Forb}(A, Y, \G_1)$, and so $W_1 \cap \mbox{Forb}(A, Y, \G_0) = \emptyset$. For the same reason, $\G_0$ can only have fewer (not more) paths than $\G_1$ and $\G_0$ can only have a subset of the colliders in $\G_1$ (no ``new'' colliders). So, if all proper non-causal paths are blocked by $W_1$ in $\G_1$, they are also blocked by $W_1$ in $\G_0$. (The result also follows from Proposition 3 in \citet{peters2015structural}.)
	\end{proof}
	
	\begin{proof}[Proof of Theorem \ref{thm:2}]
		The result follows from a modification of Lemma~4 in
		\citet{kalisch2007estimating}. Let $m_{\mathrm{reach},n}$ $=$ maximum reached value
		of $s$ (size of conditioning set) in the PC algorithm and
		$m_{n} \geq m_{\mathrm{reach},n}$. $\widehat{\mathcal{G}}_{n,\delta}(m_{n})$ refers to the
		graph estimated by equivalence-PC with reached conditioning set size
		$m_{n}$. In their Lemma~4, Kalisch and Buhlmann are concerned with two
		types of error: falsely removing some edge when it exists, and falsely
		asserting some edge when it does not exist. This theorem concerns only
		the former type of error, since the desired property is only supergraph-consistency.
		$\mathbb{P}(
		\text{ $\widehat{\mathcal{G}}_{n,\delta}(m_{n})$ incorrectly removes an edge })
		\leq \\  \mathbb{P}( \bigcup_{i,j,S} E_{ij.S}) \leq O(p^{m_{n}+2}) \sup_{i,j.S}
		\mathbb{P}(E_{ij.S})$ where $E_{ij.S}$ is the event that the test of
		$X_{i} \protect \mathpalette{\protect \independenT }{\perp }X_{j} | X_{S}$
		removes the edge between $i$, $j$ while $\rho _{ij.S} \neq 0$ (a type I error
		of the equivalence test). The leading constant follows from the bound on
		the number of tests perfomed by PC (cf. \citet{kalisch2007estimating}). The
		rejection rule is composed of two one-sided tests: consider
		$H_{0}: \rho _{ij.S} > \delta $ without loss of generality so the event
		of interest is
		$E_{ij.S} = \sqrt{n-|S|-3}
		( z(\widehat{\rho}_{ij.S}) - z(\delta ) ) < \Phi ^{-1}(\alpha _{n})
		$ while $z(\rho _{ij.S}) \neq 0$. Let
		$\alpha _{n} = (1 - \Phi (\sqrt n \delta /2))$ so
		$\Phi ^{-1}(\alpha _{n}) = - \sqrt{n} \delta /2$. Thus, we need to show \\
		$\sup_{i,j.S} \mathbb{P}   ( z(\delta) - z(\widehat{\rho}_{ij.S}) > (\sqrt n / \sqrt{n-|S|-3} ) \delta /2   )$ goes to zero.
		\begin{align*}
			&\mathbb{P} \bigl( z(\delta) - z(\widehat{\rho}_{ij.S}) > \bigl(\sqrt
			n / \sqrt{n- | S | -3} \bigr) \delta /2 \bigr)
			\\
			&\quad =\mathbb{P} \bigl( z(\rho
			_{ij.S}) - z(\widehat{\rho}_{ij.S})  > \bigl( \sqrt n / \sqrt{n- | S | -3} \bigr)
			\delta /2
			\\
			&\qquad {} + \bigl(z(\rho _{ij.S}) - z(\delta ) \bigr) \bigr) ,
		\end{align*}
		where the last term $>0$ by assumption. Then we can use Lemma~3 from
		\citet{kalisch2007estimating} to see this is quantity is bounded and goes
		to zero as $n \to \infty $.
	\end{proof}
	
	\begin{proof}[Proof of Theorem \ref{thm:3}]
		In constructing the two estimates $\widehat{\G}_{n, \delta}$ and $\widehat{\G}_{n, \delta'}$, the PC algorithm will test a series of triples $(i,j,S)$. Between the two runs, some triples will be tested in both, but the estimation of $\widehat{\G}_{n, \delta}$ may involve more tests than $\widehat{\G}_{n, \delta'}$, i.e., if a triple $(i,j,S)$ is tested with setting $\delta'$ then it is also tested with setting $\delta$, but the $\delta$ setting may involve additional tests. First, consider a triple $(i,j,S)$ that is tested in both runs. If the edge between $i,j$ is removed as a result of this test in constructing $\widehat{\G}_{n, \delta}$, it is because $|\widehat{\rho}_{ij.S}| < \delta$, but since $\delta < \delta'$ this edge would also be removed in constructing $\widehat{\G}_{n, \delta'}$. (To be precise, an edge is removed if $\sqrt{n-|S|-3} ( z(\widehat{\rho}_{ij.S}) \pm z(\delta)) < C$ for threshold $C$ but since $n$ and $C$ are fixed here and $z(\cdot)$ is monotonic we use the simpler formulation.) If $|\widehat{\rho}_{ij.S}| > \delta$, the corresponding edge is not removed in $\widehat{\G}_{n, \delta}$ and it may or may not be removed in $\widehat{\G}_{n, \delta'}$, but in any case the subgraph relation holds. Next consider a triple $(i,j,S^*)$ that is tested in constructing $\widehat{\G}_{n, \delta}$ but never tested in constructing $\widehat{\G}_{n, \delta'}$. 
		$\mathbb{P}( \sk(\widehat{\G}_{n,\delta'}) \not\subseteq \sk(\widehat{\G}_{n,\delta}))$ is equivalent to the probability that some $(i,j,S^*)$ leads to an edge removal in constructing $\widehat{\G}_{n, \delta}$ but triple $(i,j,S^*)$ is never tested and hence the $i,j$ edge is not removed in constructing $\widehat{\G}_{n, \delta'}$.
		Since the true graph either contains an edge between vertices $i,j$ or it does not, there are only two possibilities: either $\widehat{\G}_{n, \delta}$ makes an error in removing the edge or $\widehat{\G}_{n, \delta'}$ makes an error in not removing the edge.
		Regarding the first case, we have already shown in Theorem 2 that the probability of falsely removing an edge (a type I error) in constructing $\widehat{\G}_{n, \delta}$ goes to $0$ as $n \to \infty$. 
		Regarding the second case, if $\widehat{\G}_{n, \delta'}$ errs in not removing the edge between $i,j$ it must be because it never tests the corresponding correct conditioning set $S^*$. PC considers conditioning sets that are subsets of $\Adj_i(\widetilde{\G})$ where $\widetilde{\G}$ is the graph at some intermediate stage of the algorithm. If the algorithm never tests conditioning set $S^*$ in constructing $\widehat{\G}_{n, \delta'}$ but it correctly tests $S^*$ in constructing $\widehat{\G}_{n, \delta}$ then $\Adj_i(\widetilde{\G}_{n,\delta'})$ must be smaller than $\Adj_i(\widetilde{\G}_{n,\delta})$ at the corresponding stage of the algorithm, which implies that the algorithm incorrectly removed an edge adjacent to $i$ in constructing $\Adj_i(\widetilde{\G}_{n,\delta'})$. Again, the probability of falsely removing an edge $\to 0$ as $n \to \infty$.
		Therefore $\mathbb{P}( \sk(\widehat{\G}_{n,\delta'}) \subseteq \sk(\widehat{\G}_{n,\delta})) \to 1$.
	\end{proof}

\section*{Acknowledgements}

The author would like to acknowledge and thank the participants in the ``Program on Causality'' at the Simons Institute, UC Berkeley, for formative discussions related to this work. This research was partially supported by the National Institutes of Health under award number K25ES034064 from NIEHS.

\bibliographystyle{abbrvnat} 
\bibliography{../references-personal/bib}

\end{document}